# Inclusive $D^*$ Production in $\gamma\gamma$ Collisions: Including the Single-Resolved Contribution with Massive Quarks

G. Kramer[1] and H. Spiesberger[2]

[1] II. Institut für Theoretische Physik, Universität Hamburg,
Luruper Chaussee 149, D-22761 Hamburg, Germany

[2] Institut für Physik, Johannes-Gutenberg-Universität,
Staudinger Weg 7, D-55099 Mainz, Germany

**Abstract**

We have calculated the next-to-leading order cross section for the inclusive production of charm quarks as a function of the transverse momentum $p_T$ and the rapidity in two approaches using massive or massless charm quarks. For the single-resolved cross section we have derived the massless limit from the massive theory. We find that this limit differs from the genuine massless version with $\overline{\text{MS}}$ factorization by finite corrections. By adjusting subtraction terms we establish a massive theory with $\overline{\text{MS}}$ subtraction which approaches the massless theory very fast with increasing transverse momentum. With these results and including the equivalent results for the direct cross section obtained previously as well as double-resolved contributions, we calculate the inclusive $D^{*\pm}$ cross section in $\gamma\gamma$ collisions using realistic evolved non-perturbative fragmentation functions and compare with recent data from the LEP collaborations ALEPH, L3 and OPAL. We find good agreement.

# 1 Introduction

Recently there has been quite some interest in the study of charm production in two-photon collisions at high energy $e^+e^-$ colliders, both experimentally and theoretically. The three LEP collaborations ALEPH, L3 and OPAL have presented cross section data for inclusive $D^*$ production in two-photon collisions at $e^+e^-$ center-of-mass energies close to $\sqrt{s} = 189$ GeV [1, 2, 3]. Besides the total cross section for $\gamma + \gamma \to D^* + X$, also differential cross sections with respect to the $D^*$ transverse momentum, $d\sigma/dp_T$, and the pseudo-rapidity, $d\sigma/d\eta$, have been measured.

On the theoretical side two distinct approaches for next-to-leading order (NLO) calculations in perturbative QCD have been used for comparison with the experimental data. In the so-called massless scheme (ZM scheme) [4], which is the conventional parton model approach, the zero-mass parton approximation is applied also to the charm quark, although its mass $m$ is certainly much larger than $\Lambda_{QCD}$. In this approach the charm quark is also an ingoing parton originating from the photon, leading to additional single- and double-resolved contributions (besides those from $u$, $d$, $s$ quarks and the gluon $g$). The charm quark fragments into the $D^*$ meson similarly as the light quarks and the gluon with a fragmentation function (FF) known from other processes. The well-known factorization theorem then provides a straightforward procedure for order-by-order perturbative calculations. Although this approach can be used as soon as the factorization scales of initial and final state are above the starting scale of the parton distribution function (PDF) of the photon and of the FF of the $D^*$, the predictions are reliable only in the region of large transverse momenta $p_T \gg m$, where terms of the order of $m^2/p_T^2$ can safely be neglected.

The other calculational scheme which has been applied to the process $\gamma + \gamma \to D^* + X$ in [5] is the so-called massive scheme, also called fixed flavor-number scheme (FFN), in which the number of active flavors in the initial state for the resolved contributions is limited to $n_f = 3$ and the charm quark appears only in the final state of the direct, single-resolved and double-resolved contributions. In this case, the $c$ quark is always treated as a heavy particle and never as a parton. The actual mass parameter $m$ is explicitly taken into account along with $p_T$ as if they were of the same order, irrespective of their real relative magnitudes. In this scheme the charm mass acts as a cutoff for the initial- and final-state collinear singularities and sets the scale for the perturbative calculations. However, in NLO, terms $\propto \alpha_s \ln(p_T^2/m^2)$ arise from collinear emissions of a gluon by the charmed quark at large transverse momenta or from almost collinear branchings of photons or gluons into $c\bar{c}$ pairs. These terms are of order $O(1)$ for large $p_T$ and with the choice $\mu_R \sim p_T$ for the renormalization scale they spoil the convergence of the perturbation series. The FFN approach with $n_f = 3$ should thus be limited to a rather small range of $p_T \sim m$. Nevertheless, predictions in this approach have been compared to experimental data up to $p_T = 12$ GeV [1, 2].

As has been explained at many places in the literature, mostly in the context of charm production in deep inelastic $ep$ scattering (for a very recent review see [6]), the correct approach for $p_T \gg m$ is to absorb the potentially large logarithms into the charm PDF of the photon and the FF of the $c$ into $D^*$. Then, large logarithms $\propto \ln(M^2/m^2)$ defined with the factorization scale $M$ determine the evolution to higher scales and can be resummed by virtue of the Altarelli-Parisi equations. The unsubtracted terms $\propto \ln(p_T^2/M^2)$ are of order $O(1)$ for the appropriate choice $M$ of order $p_T$. After factorizing the $\ln m^2$ terms, the hard cross section is infrared safe and $n_f = 4$ is taken in the evolution equations and the equation for $\alpha_s$. The remaining dependence on $m$, i.e. the terms proportional to $m^2/p_T^2$, can be kept in the hard cross section to achieve better accuracy in the intermediate region $p_T \gtrsim m$. The factorization of mass divergent terms can be extended consistently to higher orders in $\alpha_s$, as has been shown by Collins in the context of heavy quark production in high-$Q^2$ $ep$ collisions [7].

Now it is well known that the subtraction of just the collinearly, i.e. mass, singular terms does not define a unique factorization prescription. Also finite terms must be specified. In the conventional ZM calculations the mass $m$ is put to zero from the beginning and the collinearly divergent terms are defined with the help of dimensional regularization. This fixes the finite terms in a specific way and their form is inherent to the chosen regularization procedure. If one starts with $m \neq 0$ and performs the limit $m \to 0$ afterwards, the finite terms can be different. These terms have to be removed by subtraction together with the $\ln m^2$ terms in such a way that in the limit $p_T \to \infty$ the known massless $\overline{\text{MS}}$ expressions are recovered. This requirement is actually unavoidable since all existing PDF's and FF's, including those for heavy quarks, are defined in this particular scheme (or sometimes in the DIS scheme which can be derived from the $\overline{\text{MS}}$ scheme). It is clear that a subtraction scheme defined in this way is a correct extension of the conventional zero-mass scheme to include charm mass effects in a consistent way.

In a recent work we applied this finite charm mass scheme with $\overline{\text{MS}}$ subtraction to the calculation of the cross section for $\gamma + \gamma \to D^* + X$ [8]. As a first step we considered only the direct cross section with $m \neq 0$. In the calculation of the full cross section needed for comparison with experimental data, i.e. in the sum of the direct, single-resolved and double-resolved parts, the latter two contributions were still treated in the ZM 4-flavor scheme. It is the purpose of this work to extend the finite charm mass calculation to the single-resolved cross section. This cross section plays an important part due to the partonic subprocess $\gamma + g \to c + \bar{c}$ with charm quarks in the final state and due to the process $\gamma + q \to c + \bar{c} + q$, where $q$ is one of the light (massless) quarks $q = u, d, s$. These contributions and their NLO corrections should be computed with massive charm quarks in the same way as the direct cross section due to the partonic subprocess $\gamma + \gamma \to c + \bar{c}$ and its higher order corrections. The double-resolved part with charm quarks in the final state originating from $q + \bar{q} \to c + \bar{c}$ and $g + g \to c + \bar{c}$ and the corresponding NLO corrections will still be considered in the ZM 4-flavor approach since it contributes very little to the complete double-resolved cross section [9, 5, 8].

Starting with $\gamma + g \to c + \bar{c}$, the NLO corrections for the single-resolved cross section can be split into an Abelian and a non-Abelian part. The Abelian part is, up to a constant factor, identical to the NLO corrections to $\gamma + \gamma \to c + \bar{c}$. For this part, the terms in the massive theory surviving in the limit $m \to 0$, which are not present in the ZM approach, have been identified in our earlier work [8]. Therefore, only the non-Abelian part of the NLO corrections to the photon-gluon fusion cross section and the cross section for $\gamma + q \to c + \bar{c} + q$ have to be investigated. The single-inclusive charm cross section with $m \neq 0$ has been calculated recently by Merebashvili et al. [10]. We can use these results to derive the limit $m \to 0$ and establish the subtraction terms by comparing to the $\overline{\text{MS}}$ factorized cross section derived in [11], in the same way as we did in [8] for the Abelian part. With this knowledge we can compute the finite mass corrections for the full NLO single-resolved cross section with $\overline{\text{MS}}$ factorization.

The outline of our work is as follows. In Section 2 we describe the formulae which we use to calculate the non-Abelian part of the cross section for $\gamma + g \to c(\bar{c}) + X$ and for $\gamma + q \to c(\bar{c}) + X$ with non-zero charm mass using the results of [10]. From these cross sections we derive the limit $m \to 0$ and compare with the ZM theory of [11]. The results are reported in Section 3, where also numerical tests for checking the subtraction terms are presented. Here we show how the various terms in the NLO cross section approach their corresponding massless limits for large $p_T$. After adding the already known Abelian part and the direct contribution with $\overline{\text{MS}}$ subtraction, as well as the double-resolved contribution, we compare our results to recent experimental data from LEPII in Section 4. A summary and conclusions are given in Section 5.

## 2 Calculation of the LO and NLO Differential Cross Section

The single-resolved contribution to the process $\gamma + \gamma \to D^* + X$ has many pieces. In this section we concentrate on those contributions where the charm quark appears only in the final state. We study the mass dependence in order to obtain the massless limit which is then compared with the genuine massless theory. There is only one leading-order parton process $\gamma + g \to c + \bar{c}$ with the initial gluon originating from the resolved photon. The NLO corrections to $\gamma + g \to c + \bar{c}$ are the virtual corrections and the gluonic bremsstrahlung contributions $\gamma + g \to c + \bar{c} + g$. In addition, the subprocesses $\gamma + q \to c + \bar{c} + q$ and $\gamma + \bar{q} \to c + \bar{c} + \bar{q}$, where $q$ denotes a light quark, must be taken into account in NLO. The NLO correction to $\gamma + g \to c + \bar{c}$ has an Abelian and a non-Abelian part. The Abelian part is up to a trivial factor identical to the NLO corrections for the process $\gamma + \gamma \to c + \bar{c}$, which has been considered in our previous work [8]. So, we need to calculate only the non-Abelian contribution. In the following subsection we present the LO cross section in order to fix the notation. Then we proceed to the calculation of the NLO corrections to the non-Abelian part and of the cross section for $\gamma + q(\bar{q}) \to c + \bar{c} + q(\bar{q})$.

## 2.1 Leading-Order Cross Section

We start with the process

$$\gamma(p_1) + g(p_2) \to c(p_3) + \bar{c}(p_4) + [g(k)] \qquad (1)$$

where $p_i$, $i = 1, 2, 3, 4$ and $k$ denote the momenta of the incoming photon, the incoming gluon, the outgoing $c$ and $\bar{c}$ quarks, and a possible gluon in the final state (in square brackets), which is present in the NLO corrections. We have the following invariants

$$s = (p_1 + p_2)^2, \quad t = T - m^2 = (p_1 - p_3)^2 - m^2, \quad u = U - m^2 = (p_2 - p_3)^2 - m^2 \qquad (2)$$

and

$$s_2 = S_2 - m^2 = (p_1 + p_2 - p_3)^2 - m^2 = s + t + u. \qquad (3)$$

We define the dimensionless variables $v$ and $w$ as usual by

$$v = 1 + \frac{t}{s}, \quad w = -\frac{u}{s+t} \qquad (4)$$

so that

$$t = -s(1-v), \quad u = -svw, \quad s_2 = sv(1-w). \qquad (5)$$

The leading-order cross section is

$$\frac{d\sigma_{\text{LO}}^{\gamma g}}{dvdw} = \frac{c(s)}{2}\delta(1-w)\left(\frac{t}{u} + \frac{u}{t} + 4\frac{sm^2}{tu} - 4\left(\frac{sm^2}{tu}\right)^2\right) \qquad (6)$$

where

$$c(s) = \frac{2\pi\alpha\alpha_s e_c^2}{s}. \qquad (7)$$

In (7), $e_c$ is the electric charge of the charm quark. If we compare with the leading-order cross section $d\sigma_{\text{LO}}/dvdw$ for $\gamma + \gamma \to c + \bar{c}$ in [8], we have

$$\frac{d\sigma_{\text{LO}}^{\gamma g}}{dvdw} = \kappa C_F \frac{d\sigma_{\text{LO}}}{dvdw} \qquad (8)$$

where $\kappa = \alpha_s/(8\alpha e_c^2)$. From (6) the finite charm mass corrections are clearly visible. Numerical results for the direct contribution to $\gamma + \gamma \to c/\bar{c} \to D^* + X$ can be found in our previous paper (Fig. 1 of [8]).

## 2.2 The Next-to-Leading-Order Cross Section

The NLO corrections for $\gamma + g \to c + \bar{c}$ with non-zero charm quark mass $m$ have been calculated by several groups [12, 13, 14, 10]. Only in the publication of Merebashvili et al. explicit formulae for the separate contributions due to one-loop diagrams and due to gluonic bremsstrahlung are given in a form which allows us to derive the massless limit $m \to 0$, in order to establish the subtraction terms which are needed for the cross section in the $\overline{\text{MS}}$ factorization scheme. The results in Ref. [10] are subdivided into four parts for the virtual corrections, $d\sigma_{\text{a-e}}^{\gamma g}$, $d\sigma_{\text{f}}^{\gamma g}$, $d\sigma_{\text{g}}^{\gamma g}$, and $d\sigma_{\text{h}}^{\gamma g}$, and the bremsstrahlung cross section $d\sigma_{\text{Br}}^{\gamma g}$. These separate contributions have the following structure: $d\sigma_{\text{a-e}}^{\gamma g}$ stands for the contribution of the graphs (a)–(e) in Fig. 1. It is

$$\frac{d\sigma_{\text{a-e}}^{\gamma g}}{dvdw} = \kappa C_F \frac{d\sigma_{\text{VSE}}}{dvdw} - \frac{N_C}{2}\frac{d\sigma_{\text{a-e}}}{dv}\delta(1-w) \tag{9}$$

where $d\sigma_{\text{VSE}}/dvdw$ determines the Abelian part and is given in eq. (16) of [15] and

$$\begin{aligned}
\frac{d\sigma_{\text{a-e}}}{dv} &= \frac{C(s)}{8}\left(2\tilde{A}_1\left\{2\left[\zeta_2 - \text{Li}_2\left(\frac{T}{m^2}\right)\right]\left(1+3\frac{m^2}{t}\right) - \ln\left(\frac{-t}{m^2}\right)\left(1+\frac{m^2}{t}\right) + 2\right\}\right. \\
&\left.+\tilde{A}_2\ln\left(\frac{-t}{m^2}\right) + \tilde{A}_3\left[\text{Li}_2\left(\frac{-T}{m^2}\right) - \zeta_2\right] + \tilde{A}_4 + (t \leftrightarrow u)\right)
\end{aligned} \tag{10}$$

with

$$C(s) = \frac{\alpha_s}{2\pi}c(s). \tag{11}$$

The coefficients $\tilde{A}_i$ ($i = 1, 2, 3, 4$) are functions of $m^2$, $s$, $t$ and $u$. They are given in appendix A of [10]. The result of graph (f), i.e. the contribution of the box diagram to the virtual corrections, can be written as

$$\frac{d\sigma_{\text{f}}^{\gamma g}}{dvdw} = \kappa\left(C_F - \frac{N_C}{2}\right)\frac{d\sigma_{\text{Box}}}{dvdw} \tag{12}$$

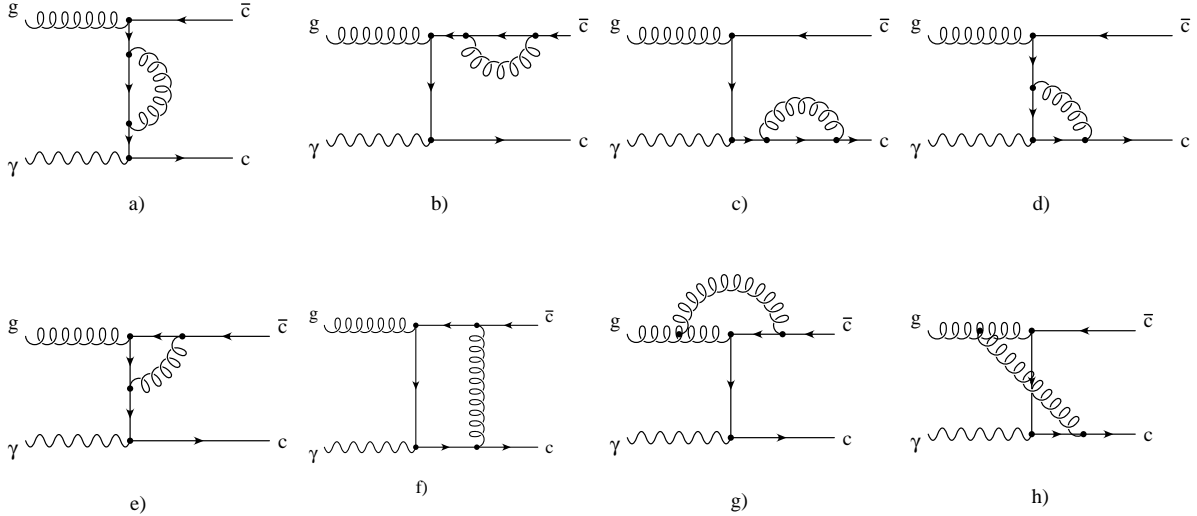

Figure 1: Feynman diagrams for the virtual NLO corrections to $\gamma + g \to c + \bar{c}$.

where $d\sigma_{\text{Box}}/dvdw$ is defined in eq. (22) of [15]. Of course, the term proportional to $C_F$ is the Abelian part and only the term proportional to $N_C$ is relevant in this work.

The diagram (g) in Fig. 1 is one of the contributions with the three-gluon coupling, which are not present in the Abelian theory. Its contribution is written as

$$\frac{d\sigma_{\text{g}}^{\gamma g}}{dvdw} = -\frac{N_C}{2}\frac{d\sigma_{\text{g}}}{dv}\delta(1-w) \tag{13}$$

where

$$\begin{aligned}\frac{d\sigma_{\text{g}}}{dv} &= \frac{C(s)}{8}\left(2A_1\left[\text{Li}_2\left(\frac{T}{m^2}\right) + \ln^2\left(\frac{-t}{m^2}\right) - 2\right] + A_1'\left[4\text{Li}_2\left(\frac{T}{m^2}\right) + 4\ln^2\left(\frac{-t}{m^2}\right)\right]\right.\\ &\quad \left. + A_2'\ln\left(\frac{-t}{m^2}\right) + A_3' + (t\leftrightarrow u)\right).\end{aligned} \tag{14}$$

In eq. (14), $A_1$ is given in appendix B of [15] and the coefficients $A_i'$, $i = 1, 2, 3$ are written in appendix A of [10].

The contribution of the non-Abelian diagram (h) in Fig. 1 is

$$\frac{d\sigma_{\text{h}}^{\gamma g}}{dvdw} = -\frac{N_C}{2}\frac{d\sigma_{\text{h}}}{dv}\delta(1-w) \tag{15}$$

where

$$\begin{aligned}\frac{d\sigma_{\text{h}}}{dv} &= \frac{C(s)}{8}\left(A_1\left[-\frac{35}{4}\zeta_2 - \text{Li}_2\left(\frac{T}{m^2}\right) + 4\ln\left(\frac{-t}{m^2}\right)\ln\left(\frac{-u}{m^2}\right) - \ln^2\left(\frac{-t}{m^2}\right)\right]\right.\\ &\quad + B_1'\text{Li}_2\left(\frac{T}{m^2}\right) + (B_2' + A_1')\zeta_2 + B_3'\ln^2\left(\frac{-t}{m^2}\right) + B_4'\ln\left(\frac{-t}{m^2}\right)\\ &\quad \left. + B_5'\ln\left(\frac{-t}{m^2}\right)\ln\left(\frac{-u}{m^2}\right) + B_6' + (t\leftrightarrow u)\right).\end{aligned} \tag{16}$$

The coefficients $B_i'$, $i = 1, 2, \ldots, 6$ are given in appendix A of [10]. In eqs. (10), (14) and (16), we left out the singular pieces in dimensional regularization proportional to $1/\epsilon^2$ ($4 - D = 2\epsilon$), and $1/\epsilon$ ($d\sigma_{\text{Box}}/dvdw$ in eq. (12) is understood without the $1/\epsilon$-term). The infrared and collinearly singular contributions cancel against the singular terms from the gluon bremsstrahlung contributions, except one remaining term which is canceled by a factorization counterterm corresponding to the final gluon emitted collinearly with the initial gluon (see Fig. 2a) which is written in the $\overline{\text{MS}}$ factorization scheme. Further details are found in [10], from which the eqs. (9) to (16) were taken.

The analytic results for the NLO corrections arising from the gluonic bremsstrahlung have also been presented in [10]. The corresponding diagrams are shown in Fig. 2a. The squared sum of the amplitudes (plus those arising from $p_1 \leftrightarrow p_2$) after summing over final spins and colors and averaging over initial colors is written as (see [10])

$$4m^2\,|M_{2\to 3}^{\gamma g}|^2 = \kappa_B(\epsilon)\left(\frac{C_F}{2}G^{\gamma\gamma} - \frac{N_C}{2}G^{\gamma g}\right) \tag{17}$$

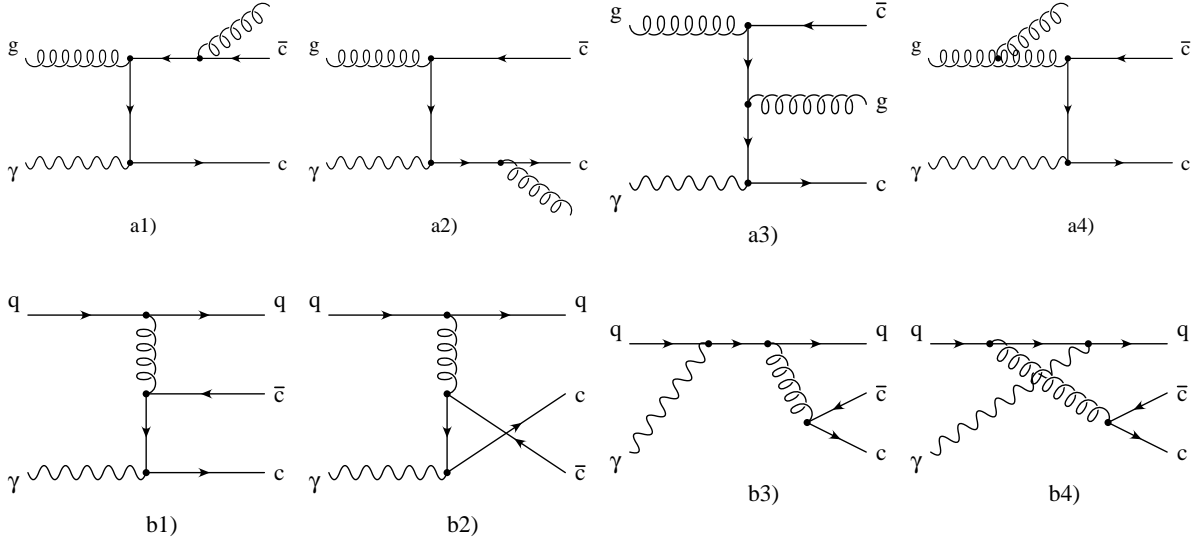

Figure 2: Feynman diagrams for the bremsstrahlung contributions $\gamma + g \to c + \bar{c} + g$ (a) and for the process $\gamma + q \to c + \bar{c} + q$ (b).

where $\kappa_B(\epsilon) = (4\pi)^3 \alpha \alpha_s^2 e_c^2 \mu^{6\epsilon}$ and $G^{\gamma\gamma}$ is the expression in the square bracket of eq. (24) of [15] (plus $p_1 \leftrightarrow p_2$). $G^{\gamma\gamma}$ has been treated already in our previous work [8]. $G^{\gamma g}$ represents the new non-Abelian part with the factor $-N_C$ instead of $C_F$. The non-Abelian bremsstrahlung contributions have been integrated also in $D \neq 4$ dimensions. The details are reported in [15] and [10]. Using $\bar{y} = \sqrt{(t+u)^2 - 4m^2 s}$ and $x = (1-\beta)/(1+\beta)$ where $\beta = \sqrt{1 - 4m^2/s}$, the final result for $d\sigma_{\text{Br}}^{\gamma g}/dvdw$ is given in eq. (4.5) of [10]. For later use we subdivide this cross section into six parts

$$\frac{d\sigma_{\text{Br}}^{\gamma g}}{dvdw} = \sum_{i=1}^{6} \left(\frac{d\sigma_{\text{Br}}^{\gamma g}}{dvdw}\right)^{(i)} \tag{18}$$

where

$$\left(\frac{d\sigma_{\text{Br}}^{\gamma g}}{dvdw}\right)^{(1)} = -\frac{C(s)N_C}{64} \frac{vs_2 s}{S_2} \left\{ e_1 + \frac{2S_2}{s_2(s+u)} e_2 \ln \frac{S_2}{m^2} + \frac{4S_2}{m^2(s+t)^2} e_3 \right.$$
$$+ \frac{2S_2}{s_2(s+t)} e_4 \ln \frac{S_2}{m^2} + e_7 I_8 + e_8 I_{10} + e_{10} I_{16}$$
$$\left. + e_{11} I_1(t \leftrightarrow u) + f_1 F_1 + f_2 F_2 \right\}; \tag{19}$$

$$\left(\frac{d\sigma_{\text{Br}}^{\gamma g}}{dvdw}\right)^{(2)} = -\frac{C(s)N_C}{64} \frac{1}{(1-w)_+} \frac{s_2^2}{S_2} \left\{ \frac{2S_2}{s_2 \bar{y}} \tilde{e}_5 \ln \frac{T+U-\bar{y}}{T+U+\bar{y}} + \tilde{e}_6 I_{11}(t \leftrightarrow u) + \tilde{e}_9 I_{11} \right\}; \tag{20}$$

$$\left(\frac{d\sigma^{\gamma g}_{\text{Br}}}{dvdw}\right)^{(3)} = -\frac{C(s)N_C}{64}\frac{vss_2}{S_2}\left\{\tilde{f}_3 F^c_3 + \tilde{f}_4 F^c_4 + \tilde{f}_7 F^c_7 + \tilde{f}_8 F^c_8 + \tilde{f}_{10} F^c_{10} + \tilde{f}_{11} F^c_{11}\right.$$
$$\left. -\left(2\ln\frac{s_2}{m^2} + \ln\frac{m^2}{S_2}\right)\left(\tilde{f}_3 F^s_3 + \tilde{f}_4 F^s_4 + \tilde{f}_8 F^s_8 + \tilde{f}_{10} F^s_{10} + \tilde{f}_{11} F^s_{11}\right)\right\}; \quad (21)$$

$$\left(\frac{d\sigma^{\gamma g}_{\text{Br}}}{dvdw}\right)^{(4)} = -\frac{C(s)N_C}{64}\frac{1}{(1-w)_+}\frac{s_2^2}{S_2}\left\{\tilde{f}_6 F^c_6 + \tilde{f}_9 F^c_9\right.$$
$$\left. -\left(2\ln\frac{sv}{m^2} + \ln\frac{m^2}{S_2}\right)\left(\tilde{f}_5 F^s_5 + \tilde{f}_6 F^s_6 + \tilde{f}_9 F^s_9\right)\right\}; \quad (22)$$

$$\left(\frac{d\sigma^{\gamma g}_{\text{Br}}}{dvdw}\right)^{(5)} = \frac{C(s)N_C}{32}\left(\frac{\ln(1-w)}{1-w}\right)_+\frac{s_2^2}{S_2}\left\{\tilde{f}_5 F^s_5 + \tilde{f}_6 F^s_6 + \tilde{f}_9 F^s_9\right\}; \quad (23)$$

$$\left(\frac{d\sigma^{\gamma g}_{\text{Br}}}{dvdw}\right)^{(6)} = \frac{\alpha_s}{2\pi}N_C\frac{d\sigma^{\gamma g}_{\text{LO}}}{dvdw}\left\{2\ln^2\frac{sv}{m^2} - \ln^2 x + \frac{1}{2}\ln^2\left(\frac{tx}{u}\right) + 2\ln\frac{u}{t}\ln\frac{sv}{m^2}\right.$$
$$+\text{Li}_2\left(1 - \frac{u}{xt}\right) - \text{Li}_2\left(1 - \frac{t}{xu}\right) - 2\zeta_2$$
$$\left. -\frac{2m^2 - s}{s\beta}\left[\left(2\ln\frac{sv}{m^2} - \ln x\right)\ln x - \text{Li}_2\left(\frac{-4\beta}{(1-\beta)^2}\right)\right]\right\}. \quad (24)$$

The coefficients $e_i$, $f_i$, $\tilde{e}_i$ and $\tilde{f}_i$ used in eqs. (19) - (24) are given in appendix B of [10]. The integrals $I_i$ and $F_i$ are found in appendix C of [15] and appendix C of [10], respectively[1].

The six pieces in $d\sigma^{\gamma g}/dvdw$ differ in the singular behavior for $w \to 1$. The second and the fourth term are proportional to $1/(1-w)_+$, whereas the fifth term contains the factor $(\ln(1-w)/(1-w))_+$. The sixth term, being proportional to $d\sigma^{\gamma g}_{\text{LO}}/dvdw$, has the factor $\delta(1-w)$. The remaining terms are finite for $w \to 1$ as long as $m \neq 0$. In the limit $m \to 0$ they give rise to additional terms proportional to $\delta(1-w)$, $1/(1-w)_+$ and $(\ln(1-w)/(1-w))_+$.

## 2.3 The Subprocess $\gamma + q \to c + \bar{c} + q$

The diagrams contributing to this process are shown in Fig. 2b. The squared sum of the amplitudes after summing and averaging over spins and colors is written as in [10]:

$$4m^2 |M^{\gamma q}_{2 \to 3}|^2 = \frac{2}{N_C}(4\pi)^3 \alpha \alpha_s^2 \left(e_c^2 Q_1 + e_q^2 Q_2 - e_c e_q Q_3\right) \quad (25)$$

where $e_q$ is the charge of the light quark $q$. The squared sum for the process $\gamma + \bar{q} \to c + \bar{c} + \bar{q}$ is given by the same expression with an opposite sign in the last term of eq. (25). The

---

[1]In the course of our work we found several misprints in Ref. [10] which are not mentioned in detail here, but can be found in [16].

result for the corresponding cross sections $d\sigma_{Q_i}^{\gamma q}/dvdw$ ($i = 1, 2, 3$) have been worked out in [10] with the following result:

$$\frac{d\sigma_{Q_1}^{\gamma q}}{dvdw} = Le_c^2 \left\{ e_1 + \frac{4S_2}{m^2(s+t)^2}e_3 + \frac{2S_2}{s_2(s+t)}e_4 \ln\frac{S_2}{m^2} \right.$$
$$\left. + e_8 I_{10} + \tilde{f}_4 F_4^c + \tilde{f}_6 F_6^c + \tilde{f}_8 F_8^c + \tilde{f}_{10} F_{10}^c + \tilde{f}_{11} F_{11}^c \right\} ; \quad (26)$$

$$\frac{d\sigma_{Q_2}^{\gamma q}}{dvdw} = Le_q^2 \left\{ e_1 + f_{12}F_{12} + f_{13}F_{13} + f_{14}F_{14} + \tilde{f}_{16}F_{16}^c + \tilde{f}_{17}F_{17}^c + \tilde{f}_{20}F_{20}^c \right\} ; \quad (27)$$

$$\frac{d\sigma_{Q_3}^{\gamma q}}{dvdw} = Le_c e_q \left\{ e_1 + \frac{2S_2}{s_2(s+t)}e_4 \ln\frac{S_2}{m^2} + e_8 I_{10} + \tilde{f}_4 F_4^c + \tilde{f}_6 F_6^c + f_{12}F_{12} + f_{14}F_{14} \right.$$
$$\left. + \tilde{f}_{16} F_{16}^c + f_{18}F_{18} + \tilde{f}_{19} F_{19}^c + \tilde{f}_{20} F_{20}^c + \tilde{f}_{21} F_{21}^c \right\} , \quad (28)$$

with
$$L = \frac{\alpha \alpha_s^2}{4N_C}\frac{vs_2}{S_2} . \quad (29)$$

The coefficients $e_i$, $f_i$, and $\tilde{f}_i$ as well as expressions for the integrals $F_i^c$ and $F_i$ are all given in appendix B and C of [10]. The integral $I_{10}$ can be obtained from appendix C of [15].

The expressions (26) and (27) do not contain terms with $1/\epsilon$ poles since they are equal with opposite sign to the corresponding counterterms. In connection with these terms there appears an additional contribution originating from the factor $(S_2 m^2/s_2^2)^\epsilon$. This yields the following contribution to the $Q_1$ part of the cross section:

$$\left(\frac{d\sigma_{Q_1}^{\gamma q}}{dvdw}\right)_{\text{eps}} = \frac{C(s)}{4} \ln\frac{s_2^2}{S_2 m^2}\frac{2sv}{1-vw}P_{gq}(x_2)B(x_2 s, t, x_2 u) \quad (30)$$

where $B(s,t,u)$ corresponds to the LO matrix element for $\gamma + g \to c + \bar{c}$ (see eq. (6))

$$B(s,t,u) = \frac{1}{s}\left[\frac{t}{u} + \frac{u}{t} + 4\frac{sm^2}{tu}\left(1 - \frac{sm^2}{tu}\right)\right] . \quad (31)$$

$x_2 = (1-v)/(1-vw)$ and $P_{gq}(x)$ is the well-known splitting function,

$$P_{gq}(x) = C_F \left(\frac{1 + (1-x)^2}{x} - \epsilon x\right) . \quad (32)$$

A similar term connected with the $1/\epsilon$ pole is present in the $Q_2$ part of the subprocess $\gamma + q \to c + \bar{c} + q$. This yields the following additional contribution:

$$\left(\frac{d\sigma_{Q_2}^{\gamma q}}{dvdw}\right)_{\text{eps}} = \frac{4C(s)}{9}\frac{e_q^2}{e_c^2} \ln\frac{s_2^2}{S_2 m^2}P_{q\gamma}(w)B_{q\bar{q}}(ws, wt, u) \quad (33)$$

where $B_{q\bar{q}}(s,t,u)$ is the LO matrix element for $q + \bar{q} \to c + \bar{c}$,

$$B_{q\bar{q}}(s,t,u) = \frac{1}{s}\left[\frac{t^2}{s^2} + \frac{u^2}{s^2} + 2\frac{m^2}{s}\right]. \tag{34}$$

The splitting function $P_{q\gamma}(x)$ is given by

$$P_{q\gamma}(x) = \frac{N_C}{2}\left[x^2 + (1-x)^2 - 2\epsilon x(1-x)\right]. \tag{35}$$

According to [10] the expressions for the $Q_1$ and $Q_2$ parts of the cross section are not complete if one needs them as derived in dimensional regularization. The results in [10] were obtained in dimensional reduction where the $\epsilon$-dependent terms of the splitting functions in eqs. (30) and (33) vanish. These terms induce extra finite terms in the limit $\epsilon \to 0$. Therefore the transition from dimensional reduction to dimensional regularization requires the addition of the following conversion terms to the $Q_1$ and $Q_2$ parts, respectively:

$$\left(\frac{d\sigma^{\gamma q}_{Q_1}}{dvdw}\right)_{\mathrm{conv}} = -\frac{C(s)}{4}\frac{2sv}{1-vw}P^{\epsilon}_{gq}(x_2)B(x_2 s, t, x_2 u) \tag{36}$$

and

$$\left(\frac{d\sigma^{\gamma q}_{Q_2}}{dvdw}\right)_{\mathrm{conv}} = -\frac{16}{9}\frac{C(s)}{4}\frac{e_q^2}{e_c^2}P^{\epsilon}_{q\gamma}(w)B_{q\bar{q}}(ws, wt, u) \tag{37}$$

where $P^{\epsilon}_{gq}$ and $P^{\epsilon}_{q\gamma}$ are the $\epsilon$-dependent parts of the splitting functions (without the factor $\epsilon$) given in (32) and (35), respectively.

The contribution $Q_3$, which arises from the interference of the diagrams in (b1) and (b2) with (b3) and (b4) in Fig. 2b, has no $1/\epsilon$ pole. Therefore there is no counterterm proportional to $e_q e_c$ and no additional contribution proportional to $\ln(s_2^2/S_2 m^2)$ like eqs. (30) and (33).

This completes the collection of formulae for the NLO corrections to $\gamma + g \to c + \bar{c}$ and the cross section for the subprocess $\gamma + q \to c + \bar{c} + q$, all based on the work of Merebashvili et al. [10] where many details must still be looked up. In order to obtain the subtraction terms for the calculation of the cross section in the $\overline{\mathrm{MS}}$ factorization scheme, we need to know the limit $m \to 0$ of the expressions given above. Of course, for the LO contribution the result is easily obtained. For the NLO corrections, however, and for the subprocess $\gamma + q \to c + \bar{c} + q$, the derivation of the limit $m \to 0$ is quite lengthy. In the next section we shall report the result of this computation and some numerical tests which we performed to check that the massless limits, from which the subtraction terms are obtained, are correct.

## 3  Zero-Mass Limit of the Massive Cross Sections

In this section we present cross sections in the limit $m \to 0$ for the NLO corrections to $\gamma + g \to c + \bar{c}$ and for the cross section $\gamma + q \to c + \bar{c} + q$ given in section 2. The result for this limit will in general be different from the cross section obtained in the approach where the mass of the charm quark is neglected from the beginning. In the genuine massless calculation, worked out in Ref. [17] and confirmed later in Ref. [11], the collinear singularities connected with the charm quark appear as $1/\epsilon$ poles in dimensional regularization. In the massive theory, they appear as terms proportional to $\ln(m^2/s)$, instead. So, in this theory the collinear singularities are regularized with a finite, although very small, charm mass. Due to this different procedure for regularizing the collinear divergent contributions, also different finite terms appear. The origin of different finite terms in these two regularization schemes lies in the fact that the two limits, $m \to 0$ and $\epsilon \to 0$, are not interchangeable. The different finite terms must be subtracted, if one wants to perform the factorization of these singular terms in the so-called $\overline{\text{MS}}$ scheme which is based on dimensional regularization with massless quarks from the start. Such finite terms have been found already in our previous work for the case of NLO corrections for $\gamma + \gamma \to c + \bar{c}$. These results for the subtraction terms are, up to a common constant $\kappa C_F$ (see eq. (8)), identical to the Abelian contributions (proportional to $C_F$) of the NLO corrections to $\gamma + g \to c + \bar{c}$. Therefore in this paper we need to establish the finite subtraction terms only for the non-Abelian part proportional to $N_C$ and for the cross section of $\gamma + q \to c + \bar{c} + q$. We write the result again in a form which has been introduced in the calculation for massless quarks by Gordon [11]. This will allow us to identify the subtraction terms we are looking for.

The LO cross section for the process $\gamma + g \to c + \bar{c}$ with $m = 0$ has the simple form (see eq. (6))

$$\lim_{m \to 0} \frac{d\sigma_{\text{LO}}^{\gamma g}}{dvdw} = \frac{c(s)}{2} \delta(1-w) \tau_0(v) \quad \text{with} \quad \tau_0(v) = \frac{v}{1-v} + \frac{1-v}{v}. \tag{38}$$

The NLO cross section and the cross section for $\gamma + q \to c + \bar{c} + q$ in the limit $m \to 0$ is decomposed in the following form:

$$\begin{aligned}
\lim_{m \to 0} \frac{d\sigma}{dvdw} &= \left(c_1 + \tilde{c}_1 \ln \frac{m^2}{s}\right) \delta(1-w) \\
&+ \left(c_2 + \tilde{c}_2 \ln \frac{m^2}{s}\right) \left(\frac{1}{1-w}\right)_+ + c_3 \left(\frac{\ln(1-w)}{1-w}\right)_+ \\
&+ c_5 \ln v + c_6 \ln(1-vw) + c_7 \ln(1-v+vw) + c_8 \ln(1-v) \\
&+ c_9 \ln w + c_{10} \ln(1-w) + c_{11} + \tilde{c}_{11} \ln \frac{m^2}{s} \\
&+ c_{12} \frac{\ln(1-v+vw)}{1-w} + c_{13} \frac{\ln w}{1-w} + c_{14} \frac{\ln(\frac{1-v}{1-vw})}{1-w}.
\end{aligned} \tag{39}$$

## 3.1 Massless Limit for NLO Corrections to $\gamma + g \to c + \bar{c}$

Here we report the results for the various coefficients $c_i$ which are written in the form

$$c_i = \hat{c}_i + \Delta c_i \tag{40}$$

where $\hat{c}_i$ are the results of [11] in the $\overline{\text{MS}}$ factorization with massless quarks and the $\Delta c_i$ are the subtraction terms we need to know.

To shorten the expressions, we make use of the abbreviations

$$X = 1 - vw, \qquad Y = 1 - v + vw, \qquad v_i = i - v. \qquad (41)$$

We shall give the result only for the non-Abelian contributions proportional to $N_C$. The Abelian terms proportional to $C_F$ are equal, up to a common factor, to the coefficients derived in our earlier work [8] and can be taken from this reference. Terms proportional to $\beta_0 = 11 - \frac{2}{3} n_f$ due to coupling constant renormalization are not shown explicitly, although they are, of course, included in our numerical evaluations. For the $c_i$ we found the following results:

$$c_1 = -C(s) \frac{N_C}{2} \left\{ 4 \ln v \ln v_1 \tau_0(v) + \ln v + \ln v_1 \right.$$
$$\left. + \left( \frac{17}{2} - \frac{7}{2v_1} - \frac{4}{v} \right) \ln^2 v + \frac{1}{2} \left( \frac{1}{v} + 1 \right) \ln^2 v_1 \right\} + \Delta c_1, \qquad (42)$$

where

$$\Delta c_1 = -C(s) \frac{N_C}{4} \frac{\ln v}{vv_1}; \qquad (43)$$

$$\tilde{c}_1 = C(s) N_C \left\{ \ln v_1 - \ln v \right\} \tau_0(v) + \Delta \tilde{c}_1, \qquad (44)$$

where

$$\Delta \tilde{c}_1 = C(s) \frac{N_C}{4} \frac{1}{vv_1}; \qquad (45)$$

$$c_2 = -C(s) N_C \left\{ \ln v_1 - 3 \ln v \right\} \tau_0(v) + \Delta c_2, \qquad (46)$$

where

$$\Delta c_2 = -C(s) \frac{N_C}{4} \frac{1}{vv_1}; \qquad (47)$$

$$\tilde{c}_2 = -C(s) N_C \tau_0(v); \qquad (48)$$

$$\tilde{c}_3 = 2 C(s) N_C \tau_0(v); \qquad (49)$$

$$c_5 = C(s)\frac{N_C}{2}\left\{-\frac{2vv_1}{X^3}+\frac{2vv_2}{X^2}-\frac{6}{v_1 X}+\frac{4v^2+6v_1}{X}+\frac{1}{w}\left[\frac{4}{v_1^2}+\frac{6}{v}-\frac{2}{v_1}-3+4v\right]\right.$$
$$\left.+w\left[\frac{8}{v_1^2}-\frac{12}{v_1}+6\right]-\frac{8}{v_1^2}+\frac{8}{v_1}-2\right\}; \tag{50}$$

$$c_6 = C(s)\frac{N_C}{2}\left\{-\frac{1}{w}\left[\frac{4}{v_1^2}+\frac{1}{v}-\frac{1}{v_1}+2\right]-w\left[\frac{8}{v_1^2}-\frac{16}{v_1}+8\right]+\frac{8}{v_1^2}-\frac{9}{v_1}+1\right\}; \tag{51}$$

$$c_7 = C(s)\frac{N_C}{2}\left\{\frac{1}{vw}+\frac{v}{v_1}\right\}; \tag{52}$$

$$c_8 = -c_6 - C(s)\frac{N_C}{2}\left\{\frac{3}{wv_1}-\frac{3}{v_1}-\frac{3}{w}+\frac{2}{vw}+3\right\}; \tag{53}$$

$$c_9 = -C(s)\frac{N_C}{2}\left\{\frac{2}{v_1 X}+\frac{2v-2-2v^2}{X}-\frac{1}{w}\left[\frac{4}{v_1^2}+\frac{2}{v}-\frac{4}{v_1}+2v+1\right]\right.$$
$$\left.-w\left[\frac{8}{v_1^2}-\frac{16}{v_1}+8\right]+\frac{8}{v_1^2}-\frac{11}{v_1}+3\right\}; \tag{54}$$

$$c_{10} = c_5 + C(s)\frac{N_C}{2}\left\{\frac{2}{v_1 X}+\frac{2v-2v^2-2}{X}+\frac{1}{w}\left[\frac{1}{v_1}-\frac{2}{v}-2v+1\right]+\frac{v}{v_1}\right\}; \tag{55}$$

$$c_{11} = C(s)\frac{N_C}{2}\left\{-\frac{2v}{Y}+\frac{6vv_1}{X^3}+\frac{v(2v-7)}{X^2}+\frac{1}{v_1 X}+\frac{v-v_1}{X}\right.$$
$$\left.-w\left(\frac{4}{v_1^2}-\frac{2}{v_1}+4\right)+\frac{4}{v_1^2}-\frac{4}{v_1}+4\right\}+\Delta c_{11}, \tag{56}$$

where

$$\Delta c_{11} = C(s)\frac{N_C}{2}\left\{\frac{2vv_1}{X^3}+\frac{vv_1}{X^2}+\frac{1}{2v_1 X}+\frac{3v-2v^2-1}{2X}-\frac{1}{2w}\left(\frac{1}{v}+2v-3\right)\right\}. \tag{57}$$

We remark that $c_{11}-\Delta c_{11}$ agrees (up to the factor $C(s)$, which is common to all $c_i$) with $\hat{c}_{11}$ in [11] for $\lambda=1$, i.e. for $D$-dimensional spin averaging.

$$\tilde{c}_{11} = -c_5 + C(s)\frac{N_C}{2}\left\{-\frac{2}{v_1 X}+\frac{2+2v^2-2v}{X}+\frac{1}{w}\left[\frac{2}{v_1}+\frac{4}{v}-4+2v\right]-\frac{4}{v_1}+4\right\}; \tag{58}$$

$$c_{12} = C(s)\frac{N_C}{2}\left\{\frac{1}{v_1}+\frac{1}{v}+2\right\}; \tag{59}$$

$$c_{13} = C(s)\frac{N_C}{2}\left\{\frac{1}{v_1} + \frac{2}{v} - 5\right\} ; \tag{60}$$

$$c_{14} = C(s)\frac{N_C}{2}\left\{\frac{1}{v} + \frac{2}{v_1} - 5\right\} . \tag{61}$$

We found subtraction terms in $c_1$, $\tilde{c}_1$, $c_2$ and $c_{11}$. $\Delta\tilde{c}_1$ is remarkable since it means that there is a term proportional to $\ln(m^2/s)$ which is not connected with a collinear divergent term as obtained with massless quarks in $\overline{\text{MS}}$ factorization. No such term was found in our previous work. The origin of $\Delta\tilde{c}_1$ remains unclear to us. In any case, its presence is not relevant for our purpose since it will be subtracted to obtain the massive theory with $\overline{\text{MS}}$ factorization. In order to check the subtraction terms $\Delta c_1$, $\Delta\tilde{c}_1$, $\Delta c_2$ and $\Delta c_{11}$, we made numerical tests which will be described in subsection 3.3.

## 3.2 Massless Limit for the Cross Section of $\gamma + q \to c + \bar{c} + q$

We shall present the cross section limits for the contributions $Q_1$, $Q_2$ and $Q_3$ separately. For all three terms there is no LO cross section of the order $O(\alpha\alpha_s)$. Therefore we have $c_1 = \tilde{c}_1 = c_2 = \tilde{c}_2 = c_3 = c_{12} = c_{13} = c_{14} = 0$ for $Q_1$, $Q_2$ and $Q_3$.

The other coefficients for $d\sigma^{\gamma g}_{Q_1}/dvdw$, which is decomposed as in (39), are:

$$c_5 = C(s)C_F\left\{\frac{v}{2X^2} - \frac{vv_3}{2Xv_1} - \frac{(1+3v^2)(1-w)}{v_1^2} + \frac{1+v^2}{v_1^2 w}\right\} ; \tag{62}$$

$$c_6 = C(s)C_F\left\{\frac{2v(1+v)}{v_1^2} - \frac{1+v^2}{v_1^2 w} - \frac{4v^2 w}{v_1^2}\right\} ; \tag{63}$$

$$c_7 = 0 ; \tag{64}$$

$$c_8 = -c_6 ; \tag{65}$$

$$c_9 = -c_6 ; \tag{66}$$

$$c_{10} = c_5 ; \tag{67}$$

$$\tilde{c}_{11} = -c_5 ; \tag{68}$$

$$c_{11} = C(s)C_F\left\{-\frac{v}{2X^2} + \frac{v}{2v_1 X} + \frac{1}{2w} - \frac{2w}{v_1^2} + \frac{2v^2}{v_1^2} + \frac{2}{v_1} + \frac{w}{v_1} - 2w\right\} . \tag{69}$$

We found no additional terms compared to [11]. Therefore in the $Q_1$ part all coefficients obtained with mass regularization agree in the limit $m \to 0$ with the massless calculation in $\overline{MS}$ factorization. We remark that the result eq. (69) for $c_{11}$ corresponds to $D$-dependent spin averaging of the photon and gluon ($\lambda = 1$ in appendix B8 of [11]). In order to obtain the results (62–69), i.e. without additional terms $\Delta c_i$, it was essential to incorporate the conversion term (36).

The coefficients for $d\sigma^{\gamma g}_{Q_2}/dvdw$ are the following:

$$c_5 = C_q(s)C_F \left\{ \frac{(3-2v)v}{2Y} - \frac{vv_1}{2Y^2} - \left(3v^2 + v_1^2\right)(1-w) + \frac{1-2vv_1}{w} \right\} ; \tag{70}$$

$$c_6 = 0 ; \tag{71}$$

$$c_7 = -C_q(s)C_F \frac{v^2 w}{Y^2} (1 - 2v + 2vw) ; \tag{72}$$

$$c_8 = 0 ; \tag{73}$$

$$c_9 = C_q(s)C_F \left\{ \frac{1-2vv_1}{w} + 2v - 4v^2(1-w) \right\} ; \tag{74}$$

$$c_{10} = c_5 ; \tag{75}$$

$$\tilde{c}_{11} = -c_5 ; \tag{76}$$

$$c_{11} = C_q(s)C_F \left\{ \frac{(3v-4)v}{2Y} + \frac{vv_1}{Y^2} + \frac{1}{w}\left(1 + v - v^2\right) \right.$$
$$\left. + w\left(3v - 1 - 4v^2\right) + 4v^2 - 2v \right\} + \Delta c_{11} , \tag{77}$$

where

$$\Delta c_{11} = -C_q(s)C_F \frac{1}{w} \left(w^2 + (1-w)^2\right) . \tag{78}$$

Here we have introduced the factor

$$C_q(s) = \frac{2\pi\alpha\alpha_s e_q^2}{s} \frac{\alpha_s}{2\pi} = \frac{e_q^2}{e_c^2} C(s) . \tag{79}$$

In the contribution $Q_2$ we found one extra term in $c_{11}$, which must be subtracted when going over to the $\overline{MS}$ factorization scheme. The result for $\tilde{c}_{11}$ agrees with the result in [11], if one compares to $\tilde{c}_{11} + \tilde{\tilde{c}}_{11}$, where $\tilde{c}_{11}$ ($\tilde{\tilde{c}}_{11}$) gives the terms originating from the collinear singularities of the initial state (final state). Since in the massive theory both have the scale $m$, we have no possibility, based on the explicit results of [10], to distinguish initial and final state. We have calculated $\Delta c_{11}$ using the result in [11] for $\lambda = 1$, i.e. with $D$-dimensional spin averaging for the photon.

The massless limit for the $Q_3$ term is obtained in the same way. The corresponding coefficients have the prefactor

$$C_{cq}(s) = \frac{2\pi \alpha \alpha_s e_q e_c}{s} \frac{\alpha_s}{2\pi} = \frac{e_q}{e_c} C(s) \tag{80}$$

and the following form:

$$c_5 = 0 ; \tag{81}$$

$$c_6 = C_{cq}(s) C_F \frac{v^2(1-2w)}{v_1} ; \tag{82}$$

$$c_7 = C_{cq}(s) C_F \frac{v^2(1-2w)}{v_1} ; \tag{83}$$

$$c_8 = C_{cq}(s) C_F 2v(1-2w) ; \tag{84}$$

$$c_9 = C_{cq}(s) C_F \left\{ \frac{3v^2}{v_1} + 2 - \frac{6v^2 w}{v_1} + \frac{8v^2 w^2}{v_1} \right\} ; \tag{85}$$

$$c_{10} = -C_{cq}(s) C_F \left\{ 2 + \frac{5v^2}{v_1} - \frac{10v^2 w}{v_1} + \frac{8v^2 w^2}{v_1} \right\} ; \tag{86}$$

$$c_{11} = C_{cq}(s) C_F \left\{ \frac{v}{Y} - \frac{v}{X} - 4w + 2 \right\} . \tag{87}$$

As expected, there are no contributions to $\tilde{c}_{11}$ since $Q_3$ does not have any collinear singularities. Since there are no $\Delta c_i$ terms, all coefficients agree with the results in [11]. In total, we have found only one $\Delta c_i$, namely $\Delta c_{11}$ in the $Q_2$ contribution. This completes the calculation of the subtraction terms needed for the $\overline{\text{MS}}$ factorization scheme with massive charm quarks.

## 3.3 Numerical Test of the Subtraction Terms

The calculation of the subtraction terms, in particular those in subsection 3.1, was rather involved. Special care had to be exercised in order to recover all the terms proportional to $\delta(1-w)$, $1/(1-w)_+$, $(\ln(1-w)/(1-w))_+$ and the remaining terms in the decomposition (39). In order to check that the $\Delta c_i$ presented in subsection 3.1 are correct, and also to see how the various contributions to $d\sigma^{\gamma g}/dvdw$ written down in subsection 2.2 behave as a function of the transverse momentum $p_T$ of the charm quark in comparison with the cross section for massless quarks, we have calculated the NLO corrections in two ways. First, we have taken the six terms from the bremsstrahlung contribution as given in eqs. (19) – (24) and the contributions of the virtual corrections $d\sigma^{\gamma g}_{\text{a-e}}$, $d\sigma^{\gamma g}_{\text{f}}$, $d\sigma^{\gamma g}_{\text{g}}$ and $d\sigma^{\gamma g}_{\text{h}}$ in eqs. (9–16) and calculated these parts as a function of $p_T$ for charm quarks with mass $m = 1.5$ GeV, and also from formulae derived directly from the expressions in subsection 2.2 in the limit $m \to 0$. We normalize these parts to the LO cross section for massless charm quarks calculated from the formula (6) in subsection 2.1 and denote them by $T_i(N_C)$:

$$T_i(N_C) = \frac{d^2\sigma^{\gamma g}_i/dydp_T}{d^2\sigma^{\gamma g}_{\text{LO}}/dydp_T(m=0)} \ . \tag{88}$$

The cross sections in the numerator and denominator of $T_i$ are actually the cross section for $e^+ + e^- \to e^+ + e^- + c + X$, i.e. the photon-photon cross section folded with both the Weizsäcker-Williams spectrum (with $\theta_{\max} = 0.033$) and the distribution of gluons inside the photon (taken from [18]). For the present purpose it is sufficient to evaluate the cross sections at rapidity $y = 0$. All other input data are chosen as in our previous work [8]: $\sqrt{s} = 193$ GeV, $\Lambda_{(n_f=4)} = 328$ MeV (i.e., $\alpha_s(m_Z) = 0.1181$). The charm quarks are not fragmented and $BR(c \to D^*) = 1$. The renormalization scale is $\mu_R = \sqrt{p_T^2 + m^2}$ and the factorization scales are equal to $M_I = M_F = m$, as inherent to the formalism in subsection 2.2. Details for how $d\sigma/dydp_T$ is calculated from $d\sigma/dvdw$ have been given for $\gamma\gamma$ scattering in eq. (45) in [8]. In this equation, the distribution function of the second photon is replaced by the convolution of the Weizsäcker-Williams spectrum with the distribution of a gluon in the photon.

The results for the ratios $T_i(N_C)$ ($i = 1, \ldots 6$), $T_{\text{a-e}}(N_C)$, $T_{\text{f}}(N_C)$, $T_{\text{g}}(N_C)$ and $T_{\text{h}}(N_C)$, are shown in Figs. 3 and 4. The dashed lines give the contribution from the massive theory ($m \neq 0$) and the full lines are the massless limit of the massive calculation. We see that the dashed curves approach the full curves for increasing $p_T$ quite nicely. This means that the extraction of additional terms proportional to $\delta(1-w)$ in the $T_i(N_C)$ which appear in the massless limit, have been extracted correctly. We observe that above $p_T = 5$ GeV the two curves, dashed and full, are already very close to each other. Larger deviations are seen only for $p_T < 5$ GeV which originate from the terms $\propto m^2/p_T^2$ we are interested in. In Fig. 5a we have plotted the sum of all the $T_i(N_C)$. We see that the results of the massive theory (dashed curve) approaches the massless limit (full curve) quite well. For $p_T > 10$ GeV the difference between the two curves is negligibly small. The dotted curve gives the result where the $\Delta c_i$ terms given in subsection 3.1 are subtracted from the massless limit.

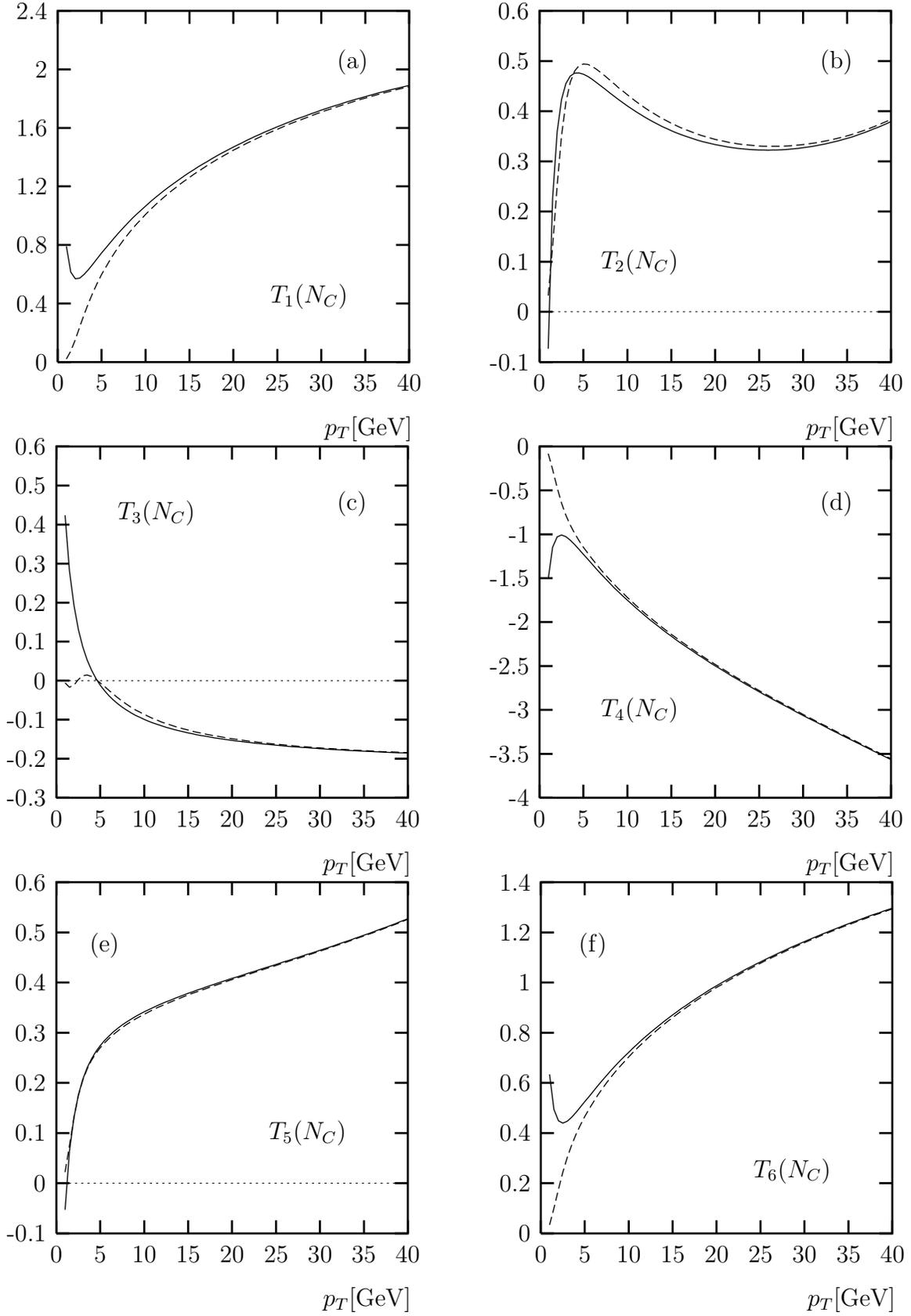

Figure 3: Comparison of the massive and massless calculations for the bremsstrahlung contributions $T_1 - T_6$ (a–f), see text.

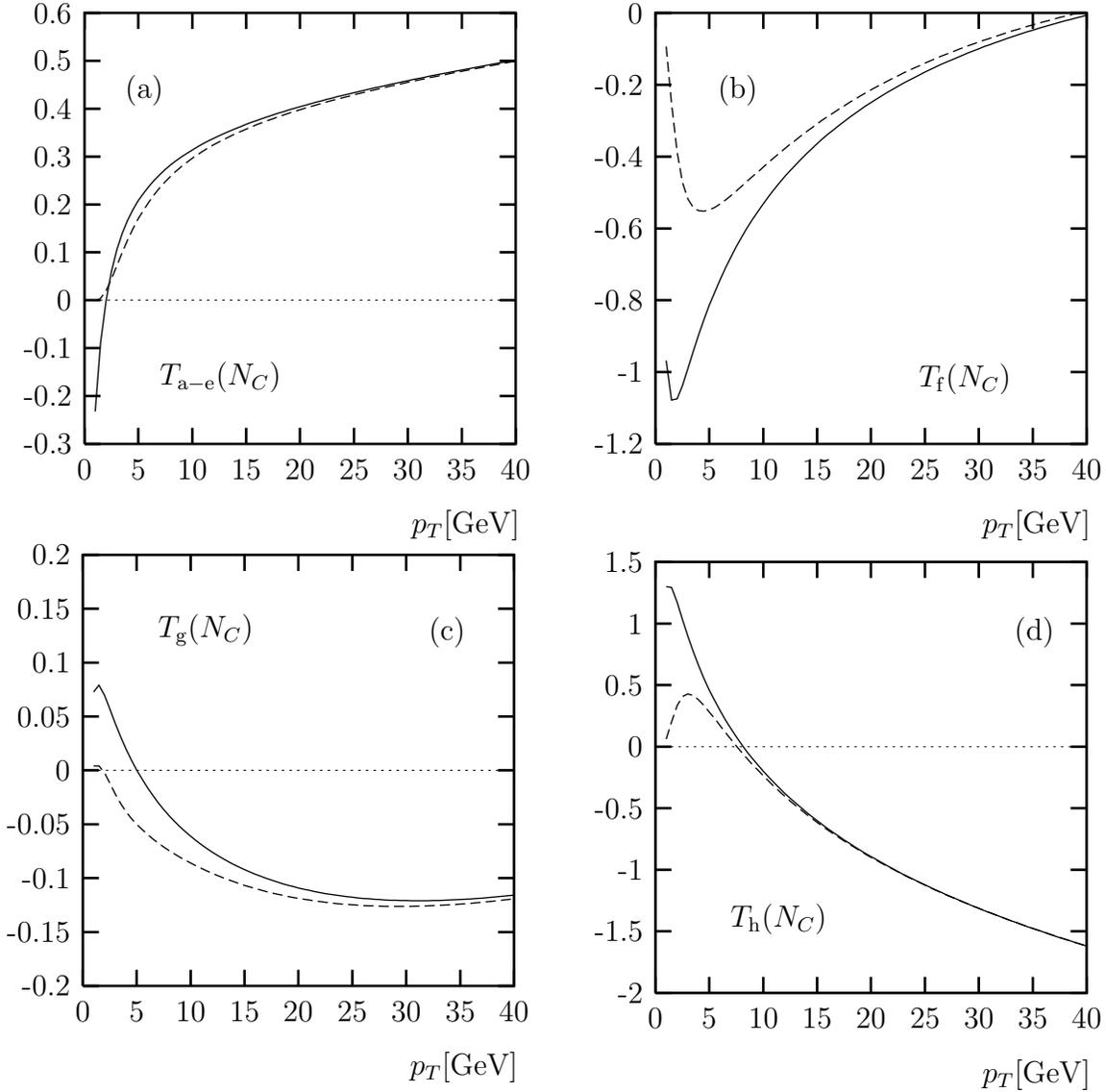

Figure 4: Comparison of the massive and massless calculations for the virtual contributions $T_{a-e} - T_h$ (a–d), see text.

This cross section is the same as the one from the massless theory with $\overline{\text{MS}}$ factorization as given in [11]. We see that the difference to the full curve is non-negligible and increases with increasing $p_T$. This difference will be added later to the massive theory in order to adjust the massive theory to the $\overline{\text{MS}}$ factorization. The difference between the massless limit and the massive cross section is more evident in Fig. 5b, where we have plotted the difference of the sum of all $T_i$ terms for the massless limit (the massless theory without $\Delta c_i$) and the massive calculation.

The sum $\sum T_i(N_C)$ in the massless limit has been calculated independently by going back to the decomposition (39) with the $c_i$ given in eqs. (42) – (61), i.e. with the subtraction terms $\Delta c_i$ included as compared to Gordon's result. For this calculation we obtained the same full curves in Figs. 5a and b. This demonstrates that the extraction of the terms $\Delta c_1$, $\Delta \tilde{c}_1$, $\Delta c_2$ and $\Delta c_{11}$ as given in subsection 3.1 is correct.

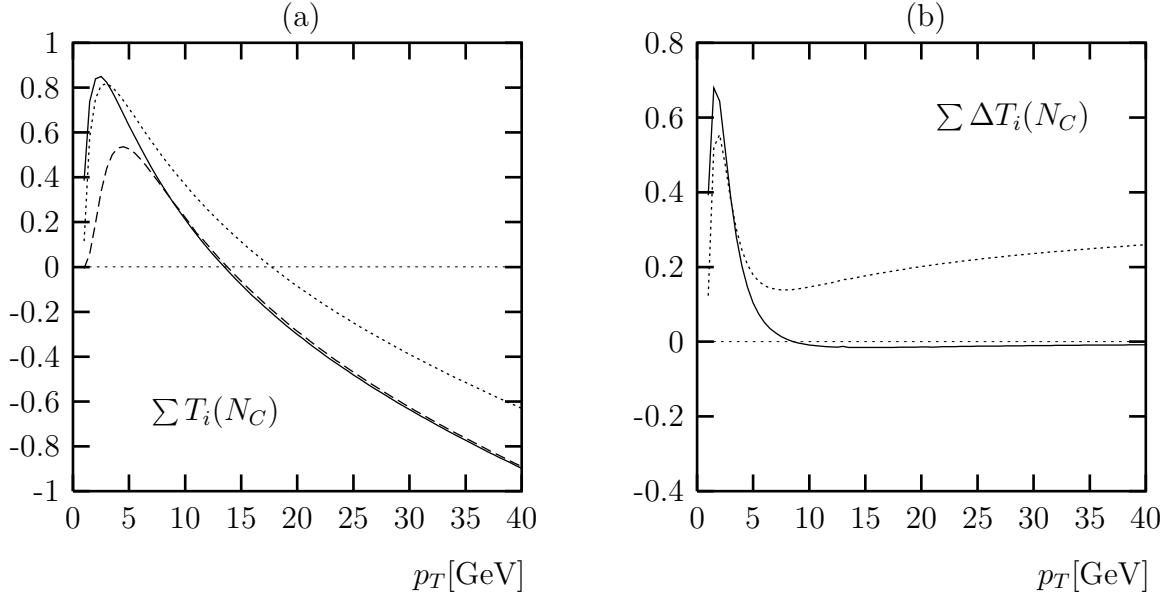

Figure 5: Comparison of the massive calculation (dashed lines) with its massless limit (full lines) for the terms proportional to $N_C$ for the process $\gamma + g \to c + \bar{c} + X$. In (a), the sum of all ratios defined in eq. (88) is shown, in (b) the difference of the sums for the massless limit and the calculation with $m \neq 0$. The dotted lines represent the results of the massless limit with subtracted $\Delta c_i$, i.e. the calculation with $\overline{\text{MS}}$ factorization [11].

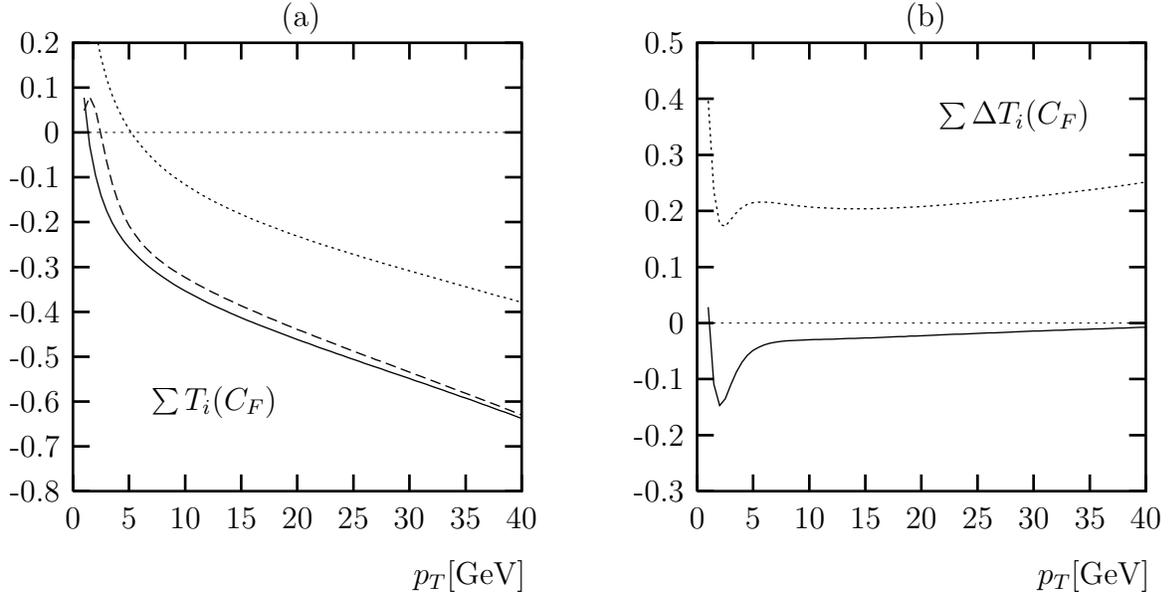

Figure 6: Comparison of massless (full lines) and massive (dashed lines) calculations for the sum of all contributions proportional to $C_F$ for the process $\gamma + g \to c + \bar{c} + X$ as in Fig. 5.

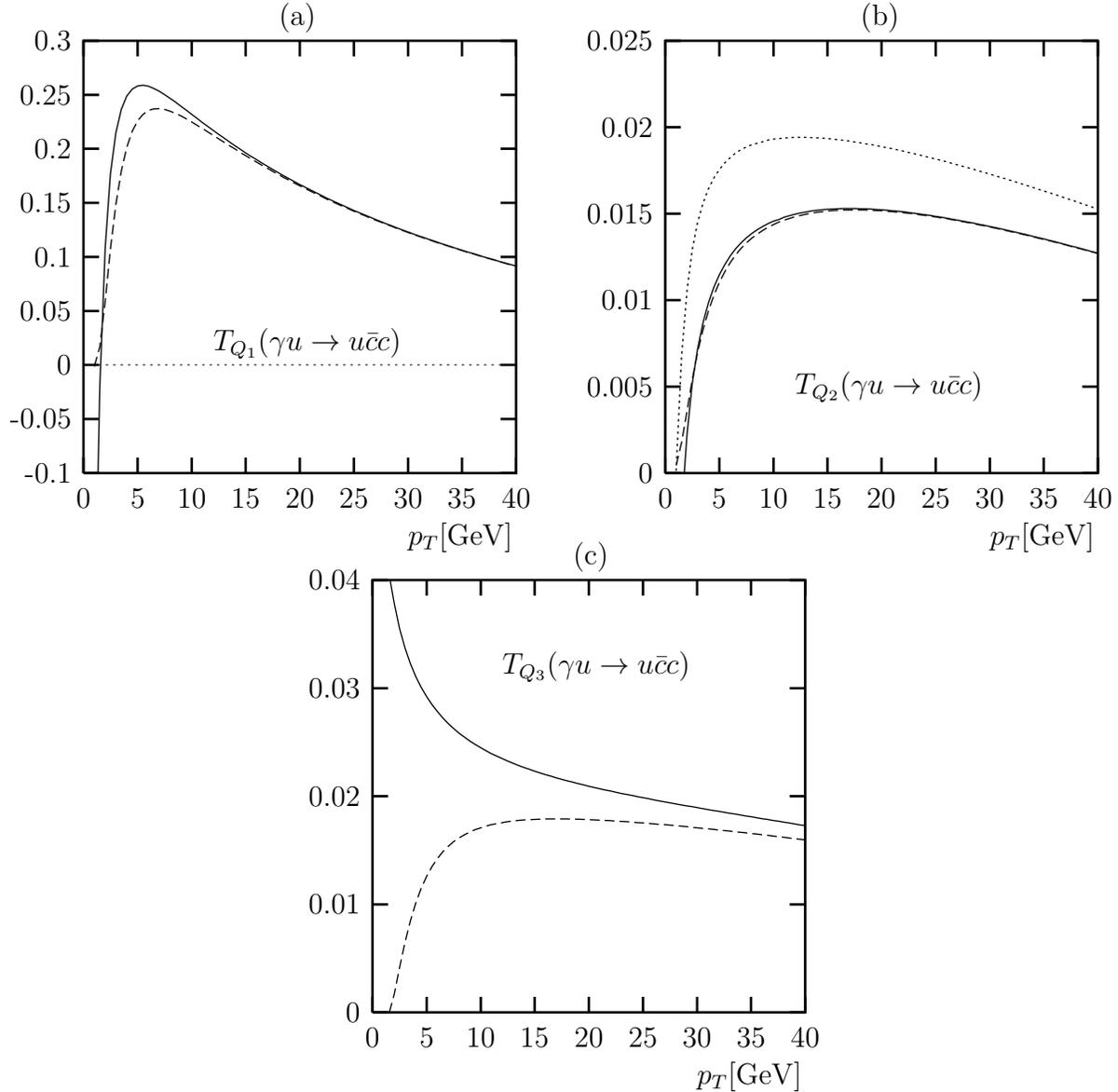

Figure 7: Comparison of the massless (full lines) and massive calculations (dashed lines) for the contributions to the process $\gamma + u \to u + c + \bar{c}$ normalized as in eq. (88). In (b) the dotted line shows the $Q_2$-part with subtracted $\Delta c_{11}$.

For comparison we have performed the same calculation for the Abelian part of the cross section. Here we have calculated $\sum T_i(C_F)$ directly from the massless limit in terms of the coefficients $c_i$, written in [8] for the process $\gamma + \gamma \to c + \bar{c}$. The results, together with the cross section for the massless theory of [11] ($\Delta c_i$ subtracted), are shown in Fig. 6a, b. We see that the massive result approaches the massless limit for large $p_T$. The difference between the two curves (see Fig. 6b) above $p_T = 5$ GeV is insignificant. We observe that the effect of the $\Delta c_i$ terms found in [8] on the $C_F$ contribution is very important.

The calculation of the subtraction term for $\gamma + q \to c + \bar{c} + q$ was much easier. So we show numerical results only for the massless limit obtained from the results in subsection 3.2. The comparison with the massive cross section for the cases $Q_1$, $Q_2$ and $Q_3$ is shown in Figs. 7a–c, again normalized to the LO cross section for $\gamma + g \to c + \bar{c}$ with $m = 0$ as in eq. (88). We see that the massless limit approaches the massive cross section quite nicely. We see also that the effect of the only $\Delta c_i$ term, namely $\Delta c_{11}$ in case of $Q_2$ (dotted curve in Fig. 7b, note the scale), has an insignificant effect.

## 4 Comparison with LEPII Data

In this section we compare our results with the experimental data from the ALEPH [1], the L3 [2] and the OPAL [3] collaborations. We do this for every experiment separately, since the three collaborations analyzed their measurements with different average center-of-mass energies $\sqrt{s}$, different anti-tagging conditions (limit on $Q^2$), and within slightly different rapidity ranges. For the rapidity distributions $d\sigma/dy$, the average over the $p_T$ range is also not identical for the three data sets. The values used for the average $\sqrt{s}$, for $Q^2_{max}$, for $|y_{max}|$ and the $p_T$ range used for the comparison are given in Tab. 1.

|  | OPAL | L3 | ALEPH |
| --- | --- | --- | --- |
| $\sqrt{s}$ [GeV] | 193 | 197 | 197 |
| tagging | $\theta < 0.033$ | $Q < m_\rho$ | $Q^2 < 4.5 \text{GeV}^2$ |
| $|y|_{max}$ | 1.5 | 1.4 | 1.5 |
| $p_T$-range [GeV] | $2 - 12$ | $1 - 12$ | $2 - 12$ |

Table 1: Experimental conditions for the measurement of inclusive $D^*$ production.

When comparing predictions for the $p_T$-distribution with the experimental results, we show both the differential cross section $d\sigma/dp_T$ as a function of $p_T$, as well as values averaged over $p_T$-bins as used by the experimental collaborations. These bins are $2 - 3$, $3 - 5$ and $5 - 12$ GeV for all three experiments and the additional bin $1 - 2$ GeV for L3. Since the $y$-distributions are very flat, there is no need to average over bins and we will show only $d\sigma/dy$.

The theoretical predictions of the massive theory consist of three parts, the direct cross section with $m \neq 0$ already presented in our previous work [8], the single-resolved contribution considered in this paper, and the double-resolved cross section. The single-resolved cross section has two components. First there are the contributions from $\gamma g \to c\bar{c}$ together with the corresponding NLO corrections and the cross section for the processes $\gamma q \to qc\bar{c}$, both with $m \neq 0$ and after subtracting the terms $\Delta c_i$ as given in sect. 3 and in our earlier work for the Abelian part [8]. Second, we add the contributions coming from the charm PDF in the photon. This part is evaluated with $m = 0$ in the hard scattering cross section including NLO corrections, using the formulae given in Ref. [11].

The particular prescription for the treatment of the incoming heavy charm quark in the single-resolved contribution as a massless parton is in fact mandatory. Since all available charm PDF's are determined with $m = 0$ in the hard scattering cross section, this is the only consistent choice. The finite charm quark mass appears only in the starting scale $\mu_0 = m$ with the effect that the charm PDF vanishes below the scale $\mu_0$ [18]. This prescription is usually also applied in the treatment of heavy quarks in deeply inelastic scattering, as advocated in [7, 19, 6].

We will also compare with predictions of the massless theory. In this case, all contributions are calculated with $m = 0$, again using the results of Ref. [11]. In both the massless and the massive calculations, we add the double-resolved contributions evaluated with $m = 0$. The corresponding formulae are based on the work of [20].

In the following the fragmentation $c \to D^*$ is calculated with the purely non-perturbative FF of Binnewies et al. [4] (OPAL set at NLO) determined from OPAL $e^+e^- \to D^*X$ data as explained in detail in [4]. We choose the renormalization scale and $\alpha_s(m_Z)$ as stated above, include $n_f = 4$ flavors and choose the factorization scales $M_I = M_F = 2\sqrt{p_T^2 + m^2}$. The transition from $M_I = M_F = m$ to this choice of scales is performed, as described in our previous work (see eq. (42) in [8]), using the coefficients $\tilde{c}_1$, $\tilde{c}_2$ and $\tilde{c}_{11}$, i.e. based on the massless calculation. The coefficients of the massless calculation must be used here, since the evolution of PDF's and FF's is also based on massless evolution kernels. This choice of scales allows us to calculate $d\sigma/dp_T$ down to small $p_T$, since otherwise we would come below the starting scale of the non-perturbative FF of [4], which is approximately equal to $2m$. We have calculated $d^2\sigma/dydp_T$, integrated over the $y$ or $p_T$ ranges as given in Tab. 1, for the three data sets. $y$ is the pseudo-rapidity as used in the analysis of the experimental data [1, 2, 3]. We identify in our calculations the pseudo-rapidity of the $D^*$ with the rapidity of the charm quark. Finally, we note that we neglect the small contribution from the fragmentation of gluons, $g \to D^*$.

In order to show the amount originating from the charm content in the photon, we have plotted $d\sigma/dp_T$ as a function of $p_T$ for the massive single-resolved contribution, i.e. without the charm contribution in the photon, and the sum of both contributions in Fig. 8a. We see that for small $p_T$ the component due to light quarks and gluons in the initial state is dominant. For large $p_T$ the contribution of the charm PDF increases, as to be expected, and amounts to 68 % of the sum at $p_T = 12$ GeV. It is clear that due to this component the influence of the charm quark mass, i.e. the correction from the $m^2/p_T^2$ terms, diminishes. This is shown in Fig. 8b, where the total single-resolved cross section is plotted for the massive and the massless calculation. On this logarithmic plot, the influence of the finite charm mass is visible only for small $p_T < 2$ GeV. At $p_T = 2$ GeV, the $m \neq 0$ cross section is reduced by $\sim 16$ % as compared to the massless approximation, as is seen more clearly in Fig. 8c where the ratio is shown. The dashed curve in this plot shows the ratio before the charm in the photon contribution is added. The strong increase of this ratio for $p_T < 2$ GeV is caused by the large NLO corrections in the massless cross section. For definiteness, the cross sections in Fig. 8 have been calculated with the ALEPH kinematical constraints (see Tab. 1).

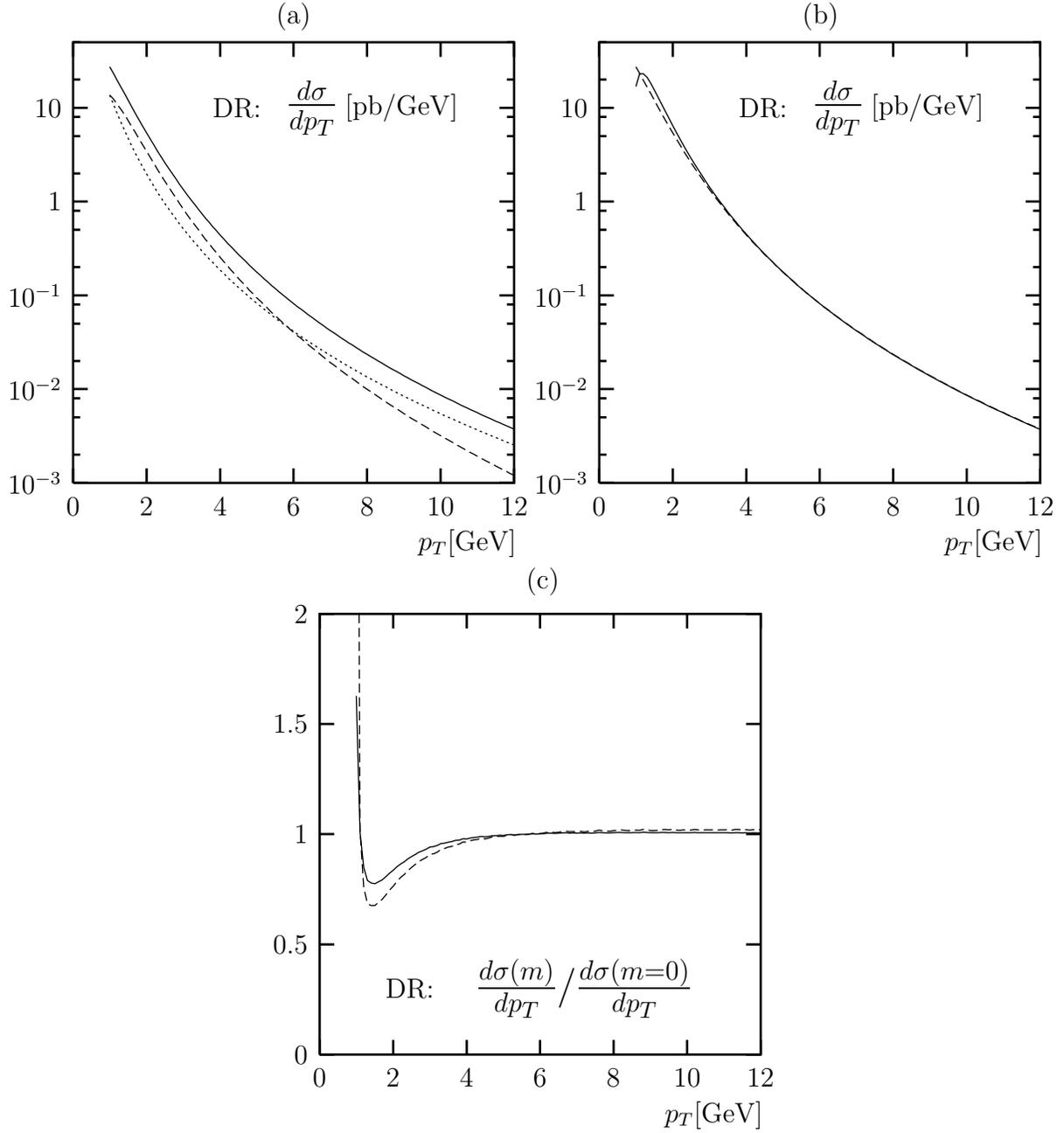

Figure 8: The single-resolved contribution: (a) shows the charm-initiated contribution (dotted line), the contribution due to gluons and light quarks (dashed line) and the sum (full line); (b) compares the sum of all contributions for the massless (full line) and the massive (dashed line) calculation; (c) shows the ratio of the massive over massless calculation with (full) and without (dashed line) charm in the photon.

The influence of the $m^2/p_T^2$ terms is further diminished in the small $p_T$ region, if we add the double-resolved cross section. This part is almost entirely due to the charm component in the photon, which we have calculated with $m = 0$ in the hard scattering parton-parton cross sections in the LO and NLO terms in the same way as we have done for the corresponding single-resolved cross section. To demonstrate this, we show in Fig. 9 the double-resolved cross section $d\sigma/dp_T$ due to light quarks and gluons in the initial state, i.e. without the charm component, together with the sum of all contributions. We see that the cross section without charm in the initial state is indeed very small in its absolute value and the double-resolved cross section is dominated by the charm component. The light-quark plus gluon component becomes negative for $p_T < 7$ GeV due to NLO corrections. This negative contribution leads to an insignificant decrease of the total double-resolved cross section for $p_T > 2$ GeV. At LO, the cross section without charm in the photon is essentially given by the $gg \to c\bar{c}$ component, which is small, since the gluon component in the photon PDF is not large. Another contribution at LO is from $qq \to c\bar{c}$, which also is expected to be small. These non-charm contributions are computed in the approximation

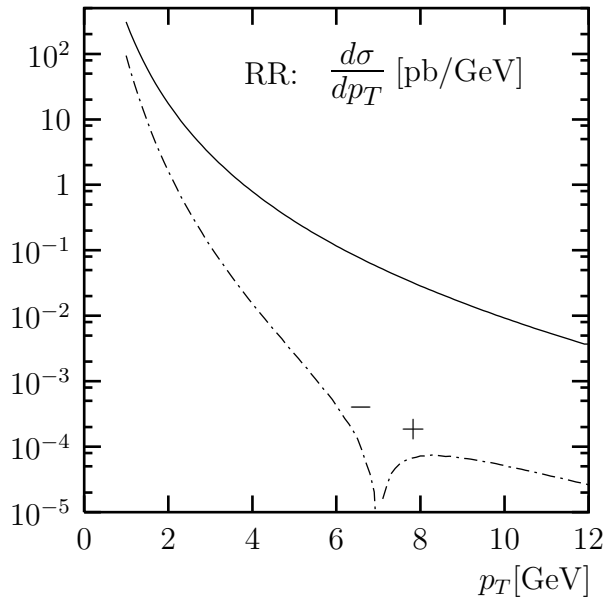

Figure 9: The double-resolved contribution. The contribution due to light quarks and gluons in the initial state is shown by the dash-dotted line. This part is negative at small $p_T$ ($p_T < 7$ GeV) where the absolute value is shown.

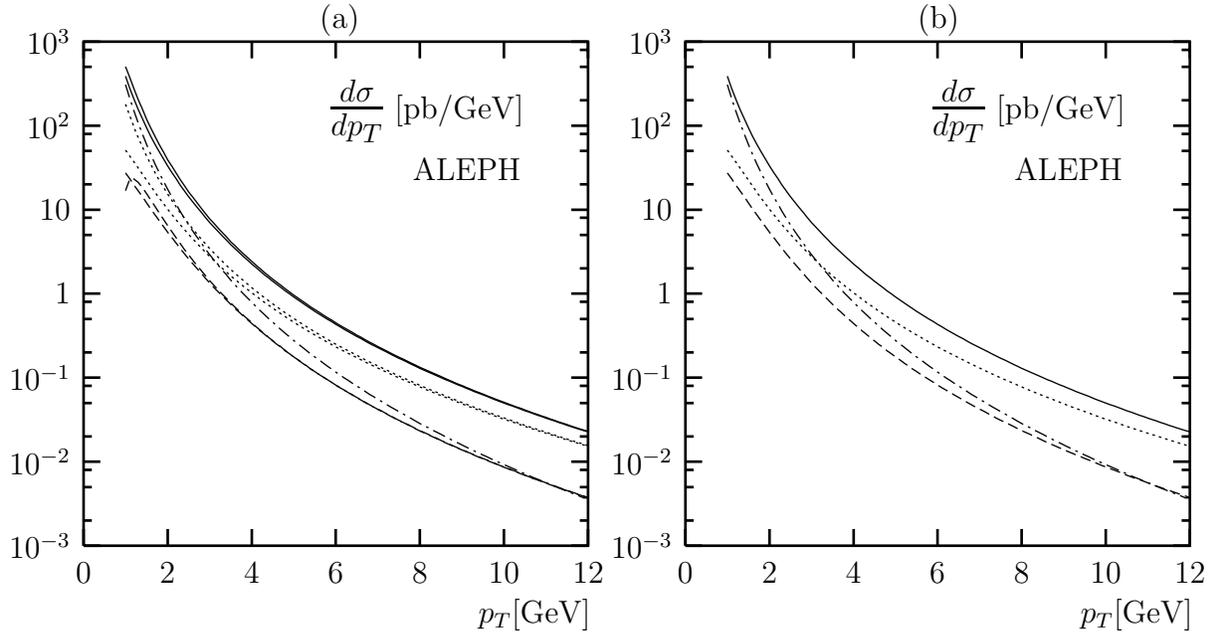

Figure 10: The partition of the full calculation (full lines) into direct (dotted), single-resolved (dashed) and double-resolved (dash-dotted) contributions. In (a), upper curves correspond to the massless and lower curves to the massive calculation. The results of the massive calculation are shown separately in (b) again for clarity.

with $m = 0$. The fact that their contribution is negligible is also true for $m \neq 0$ [9, 5].

To obtain an overview of the partition of $d\sigma/dp_T$ into direct and resolved contributions and the influence of the massless approximation, we show in Fig. 10a these cross sections with cuts as in the ALEPH analysis. All three cross sections are calculated up to NLO. At $p_T = 2$ GeV, the three contributions, direct, single-resolved and double-resolved, amount to approximately 39 %, 16 % and 45 % of the total sum, whereas at $p_T = 12$ GeV the relative contributions are 68 %, 16 % and 16 %, respectively. These numbers refer to the cross section with $m \neq 0$ except for the contributions with charm in the photon PDF. With increasing $p_T$, the resolved cross sections decrease much stronger than the direct component. In Fig. 10a we also show the direct and single-resolved cross sections for $m \neq 0$ and for $m = 0$ for comparison. We see that the finite charm mass effects, i.e. the $m^2/p_T^2$ terms, are essential only for small $p_T \leq 2$ GeV and are largest for the direct component. They lead to a reduction of the direct cross section. With this reduction the double-resolved cross section becomes dominant and therefore the influence of the finite charm mass effects is very much reduced in the total sum. For a clearer presentation of the three components in the cross section with $m \neq 0$, we show them in Fig. 10b again, where the dominance of the double-resolved part for small $p_T$ is clearly visible.

After these preparatory studies we are in the position to compare our results with the data of the three LEP collaborations. This is shown in Figs. 11a, b, c where we compare our results of the calculation with massive ($m \neq 0$) and massless ($m = 0$) charm quarks with the experimental data of the ALEPH [1], L3 [2] and OPAL [3][2] collaborations. The data are always averages over the respective bins in $p_T$: $1-2$ (L3), $2-3$, $3-5$, $5-12$ GeV (ALEPH, L3 and OPAL), but we show theoretical predictions for both the differential cross section $d\sigma/dp_T$ as well as for corresponding bin averages. The agreement between data and theory is quite satisfactory on average. The L3 point in the first bin is lower than the theoretical prediction, the results in the bin $2-3$ GeV are in good agreement for all experiments, while our results for the last two bins tend to be lower than the measured values. Only the L3 point in the highest $p_T$-bin is again in perfect agreement with the predictions. In Figs. 11a, b, c we have plotted also the pure $m=0$ cross sections. The results are not very different from the massive cross sections, even at low $p_T$. This is due to the large contribution of the double-resolved cross section at small $p_T$, which is there of the same order as the direct contribution. In general, the massive theory is below the pure massless theory. But the reduction is not very large and converges to zero with increasing $p_T$ by construction. In Fig. 11d we have collected all three data sets in one plot and compare them with the respective theoretical predictions. We see that the theoretical cross sections with ALEPH and OPAL constraints almost coincide. Only the L3 prediction is lower due to the different rapidity cut and the different anti-tagging condition. We remark again that we have in general good agreement for $p_T \geq 2$ GeV. Below this $p_T$ value, the theoretical prediction lies higher by an appreciable amount as compared to the L3 data point.

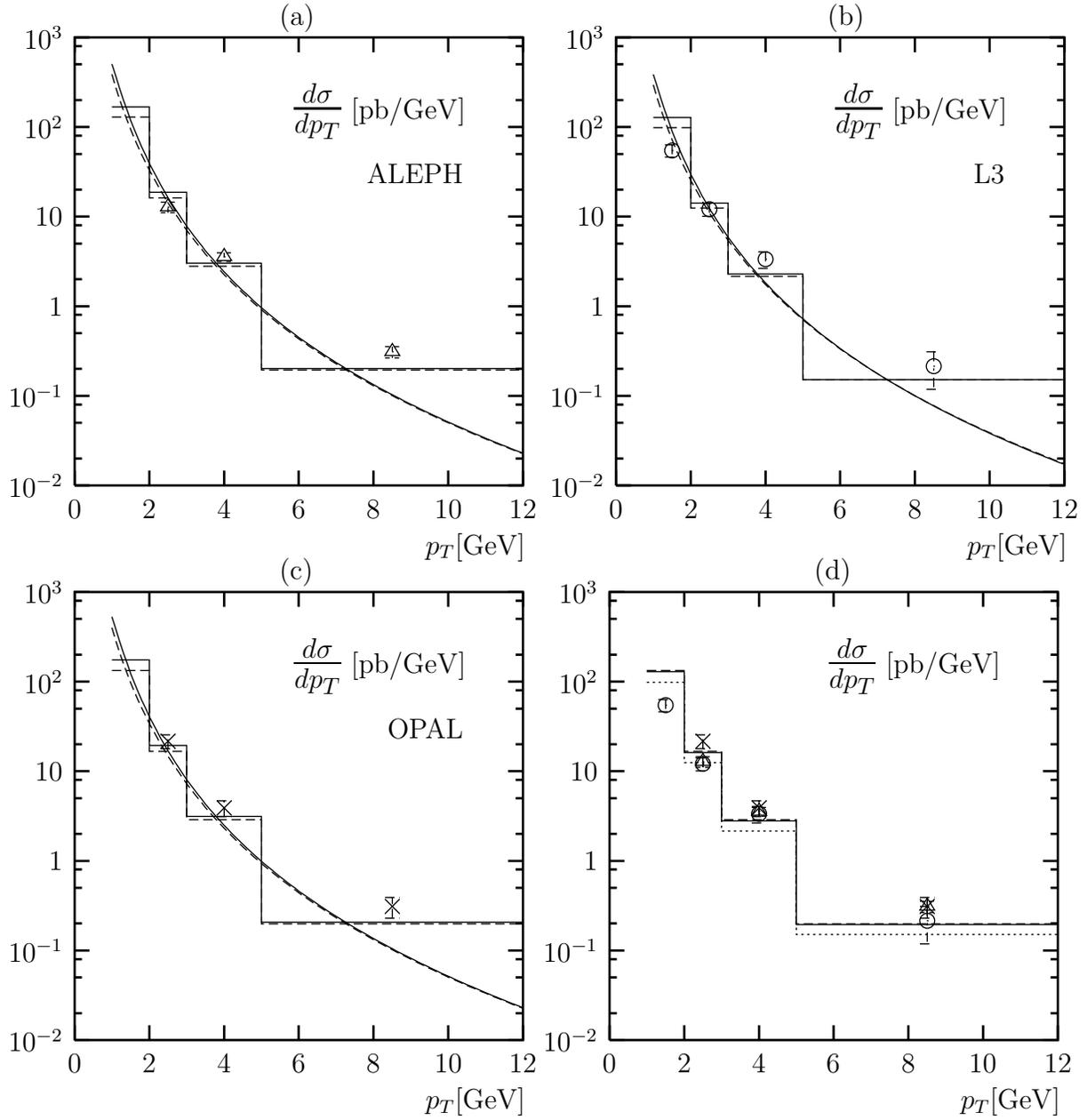

Figure 11: The full calculation compared with experimental data from ALEPH (a), L3 (b) and OPAL (c). Full lines represent the massless, dashed lines the massive calculation. Histograms show $d\sigma/dp_T$ averaged over the corresponding $p_T$-bins. The histograms in (d) are for the massive calculation (full line: ALEPH, dashed line: OPAL, dotted line: L3).

---

[2]We take the OPAL data from the second reference in [3].

Since the double-resolved cross section is dominated by the contribution due to the charm content in the photon, which is evaluated with $m=0$ for consistency as explained earlier, this part is divergent in the limit $p_T \to 0$. This explains the strong increase towards small $p_T$. This strong increase in the small $p_T$ region could be eliminated by switching to the fixed flavor theory with $n_f = 3$ below some $p_T$ value. This would eliminate the resolved contribution completely and the whole cross section would be given by the contributions with charm quarks only in the final state and not in the initial state. In the three-flavor theory, non-perturbative effects from the charm distribution in the photon, and possibly from the fragmentation $c \to D^*$, would not be present. A similar procedure has also been proposed in many papers on charm production in DIS (see for example [6]). The problem, however, lies in the fact that the precise position of the matching point is unknown and, secondly, that there is not a unique presription for how to achieve a continuous matching. The matching point will certainly lie somewhere at small $p_T$, say at $p_T \leq 2$ GeV. On the other hand, the smallest possible value of $p_T$, where the matching must occur, is, of course, the point where the photon PDF or the charm FF vanishes. With the factorization scale considered so far, this is $p_T = 0$. It is obvious therefore that these naive considerations do not provide a criterium which can be used to fix the matching scale. We might come back to this question in a future study.

The partition of the cross section into the three separate parts depends on the choice of the factorization scales. For example, for $M_I = M_F = \sqrt{p_T^2 + m^2}$ the direct contribution is larger by approximately $10\,\%$ for all $p_T$ values, and the single-resolved contribution becomes steeper (with an increase of $38\,\%$ at $p_T = 2$ GeV and $10\,\%$ at 12 GeV). In contrast, the double-resolved contribution is reduced (by $60\,\%$ at $p_T = 2$ GeV and $34\,\%$ at $p_T = 12$ GeV). Of course, the change of these separate parts is unphysical; only the change of the sum is relevant. The cross section $d\sigma/dp_T$ for the separate pieces and their sum with the new scale choice is shown in Fig. 12a (for the ALEPH experimental setup) which can directly be compared with Fig. 10b, where the results were given with the scale $2\sqrt{p_T^2 + m^2}$. This specific variation of the factorization scales amounts to a 22 % decrease at $p_T = 2$ GeV and a 1 % increase at $p_T = 12$ GeV of the physical cross section. As a consequence, the agreement of theory and data in the first bin improves with the new scale, but worsens slightly in the second bin (see Fig. 12b). One should keep in mind that continuous variations of the renormalization and factorization scales would lead to an error band of theoretical predictions which, however, we do not show.

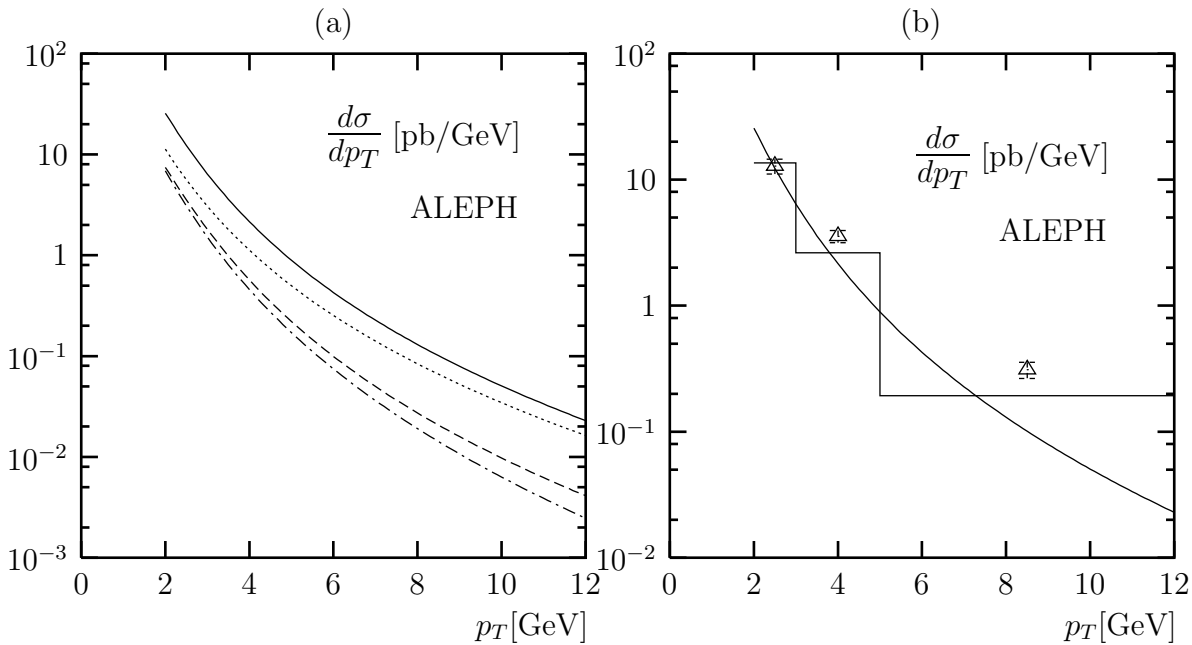

Figure 12: Effect of varied factorization scales: $M_I = M_F = \sqrt{p_T^2 + m^2}$ is chosen in this figure. (a) should be compared to Fig. 10b, (b) to Fig. 11a.

The LEP collaborations have measured also the cross section $d\sigma/dy$ as a function of $y$, where $d^2\sigma/dydp_T$ is integrated over the $p_T$ regions $2 - 12$ GeV (ALEPH and OPAL) and $1 - 12$ GeV (L3). We have calculated these rapidity distributions and compare them with the respective data points from ALEPH, L3 and OPAL in Figs. 13a, b, c. The ALEPH and OPAL points agree with the theoretical prediction inside the experimental errors (the OPAL points are slightly above the $m \neq 0$ prediction). The L3 points, however, lie below the theoretical curves. Since the cross section after integration over $p_T$ is dominated by the contribution from the lowest $p_T$ bin, this is consistent with the comparison of the $p_T$ distribution shown earlier at the smallest $p_T$ bin. In the three figures we also show how the total $y$-distribution is separated into the direct, single-resolved and double-resolved parts for both the $m \neq 0$ and the $m = 0$ theory. In Fig. 13b one can see clearly that the cross section with $m = 0$ (the upper full line) is not a reliable approximation if integrated down to $p_{Tmin} = 1$ GeV. The cross sections shown in Figs. 13a, b, c are again calculated with the scale $2\sqrt{p_T^2 + m^2}$.

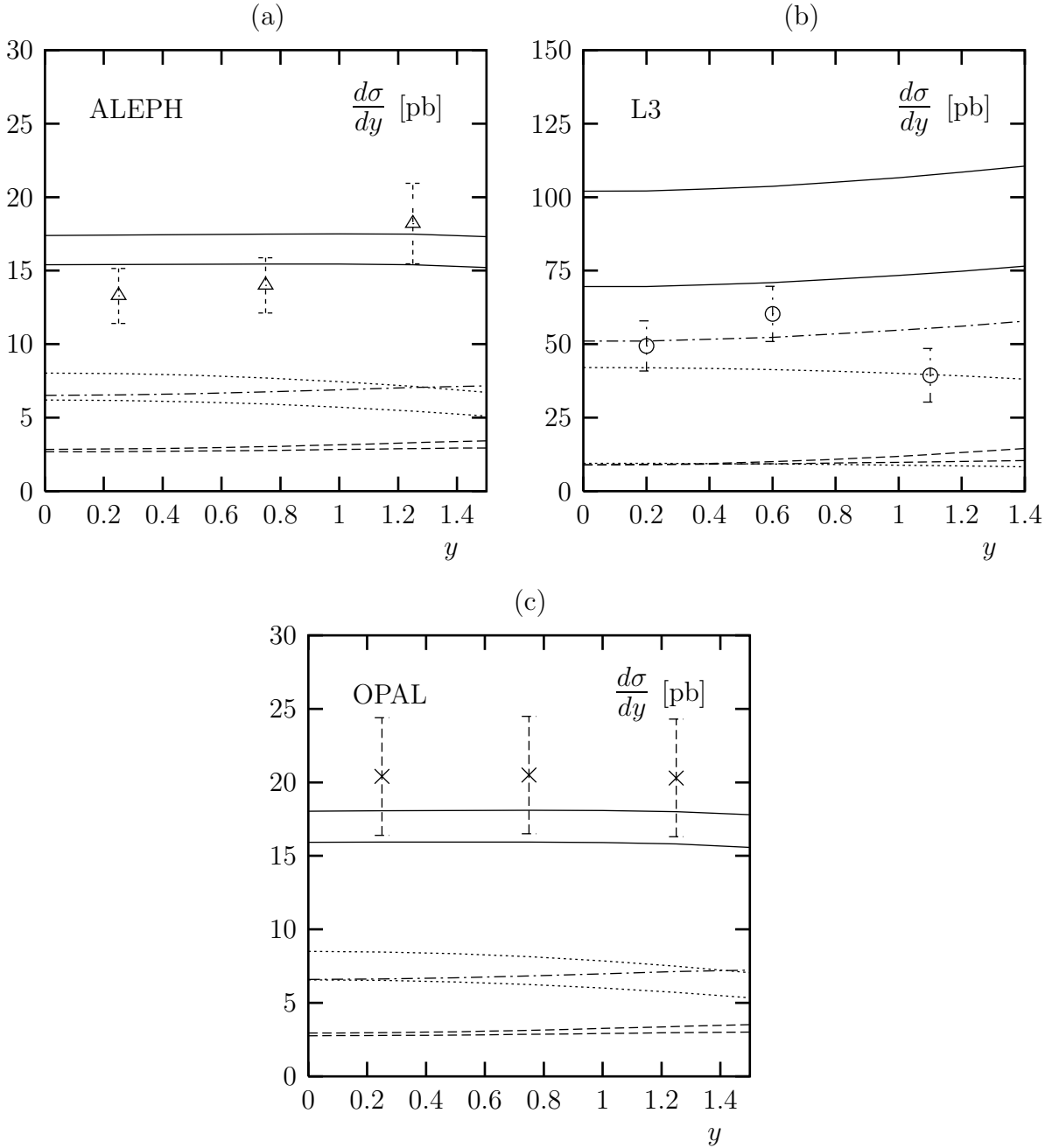

Figure 13: Theoretical predictions and experimental results for the $y$-distribution ((a): ALEPH, (b): L3, (c): OPAL). We show separately the direct (dotted), single-resolved (dashed) and double-resolved (dash-dotted) contributions and their sums (full lines). Upper curves are for the massless, lower curves for the massive calculation.

# 5 Summary and Conclusions

In this work we have compared two approaches for the calculation of inclusive charm production. One is based on a calculation with massless quarks and $\overline{\text{MS}}$ factorization, the second on a calculation with massive charm quarks. By considering the massless limit of the massive calculation, we were able to derive subtraction terms which allowed us to combine the massive calculation in a consistent way with parton distribution and fragmentation functions defined in the $\overline{\text{MS}}$ factorization scheme.

The cross section for the direct component of the $\gamma\gamma$ reaction in NLO was studied already in a previous work. Here we extend the study to the single-resolved contributions in NLO. The NLO corrections to $\gamma g \to c\bar{c}$ consist of an Abelian part and a non-Abelian part. The first part is identical, up to a normalization factor, to the NLO corrections for the direct contribution. For the second, the non-Abelian part, we found also that the massless limit of the massive cross section differs from the massless theory with $\overline{\text{MS}}$ factorization by finite terms which are non-singular for $m \to 0$. These finite terms must be subtracted from the massive hard cross section, since the latter has to be folded with a fragmentation function for the transition $c \to D^*$ and with parton distribution functions of the photon which are available only in the $\overline{\text{MS}}$ factorization scheme based on calculations with massless quarks.

It turns out that this massive theory with $\overline{\text{MS}}$ factorization leads to cross sections which converge rapidly to their massless limits with increasing $p_T$. Only at rather small $p_T$, terms proportional to $m^2/p_T^2$ are important. These terms are more important for the direct cross section than for the single-resolved one, since the latter receives contributions also from the charm distribution in the photon which must be evaluated with zero charm mass.

The double-resolved contribution has also two parts. One originates from light quarks and gluons in the initial state, the other is due to an initial charm quark in one of the two scattering photons. This latter contribution overwhelms the double-resolved cross section by far and is computed with massless quarks. Since the part with charm quarks in the final state, on the other hand, is negligible except possibly for very small $p_T$, the total double-resolved cross section can safely be evaluated with massless charm quarks.

For reliable predictions one needs a good description of the fragmentation process. In our numerical evaluation of the inclusive $D^*$ production we have taken the fragmentation functions from fits to $D^*$ production in $e^+e^-$ annihilation at LEP1. To compare with recent measurements of the $p_T$ and $y$ distributions in $\gamma\gamma$ collisions at LEP2, we added the direct, single-resolved and double-resolved contributions. The agreement of our predictions with the data is quite good (see Fig. 11) down to $p_T \simeq 2$ GeV. Finite charm mass effects are essential only for values of $p_T$ below 3 GeV. To improve the theory at very small $p_T$ it seems necessary to switch from the four-flavor theory to the three-flavor theory at some matching point.

# Acknowledgement

We thank Z. Merebashvili for clarifying communications about Refs. [10, 16] and B.A. Kniehl and M. Spira for providing us with programs for the calculation of the double-resolved contributions.